\documentclass[aps,prd,reprint,superscriptaddress,floatfix]{revtex4-1}

\usepackage[colorlinks,allcolors=blue]{hyperref}

\usepackage{graphicx}
\usepackage{subcaption}
\usepackage{rotating}
\usepackage{multirow}
\usepackage{makecell} 

\usepackage{amsmath}
\usepackage{xcolor}
\definecolor{highlight1}{HTML}{98D4F1}
\definecolor{highlight2}{HTML}{E9B08B}

\usepackage{afterpage}

\newcommand*\diff{\mathop{}\!\mathrm{d}}

\usepackage{xspace} 

\begin{document} 

\title{Interaction-model dependence in calorimetric energy reconstruction methods\\
due to non-linear material effects in modern neutrino detectors}

\author{Katharina Lachner} 
    \email{klachner@ethz.ch} 
    \affiliation{University of Warwick}
    \affiliation{ETH Zurich} 
\author{Steven Boyd} 
    \affiliation{University of Warwick} 
\author{Stephen Dolan}
    \affiliation{CERN} 
\author{Laura Munteanu} 
    \affiliation{CERN} 

\begin{abstract} 
    
Neutrino oscillation experiments rely on high precision neutrino energy reconstruction. A common reconstruction technique in LAr-TPCs and scintillators is via the calorimetric sum of visible particles created in the interaction. However, non-linearities
in the detector response, such as Birks quenching for scintillators and recombination effects for TPCs, lead to ambiguities in the reconstruction of visible hadronic energies. Interaction-model-dependent assumptions are required to resolve these ambiguities, which introduces a bias in reconstruction of neutrino energy. This introduces a systematic uncertainty separate from the well-studied bias due to interaction-model dependent modelling of missing energy caused by, for example, the production of final state neutrons. In this work, we evaluate the interaction-model dependence of the bias caused by these material effects across multiple tunes of the GENIE, NEUT, NuWro, and GiBUU neutrino interaction event generators for cases representative of calorimetric energy reconstruction at the T2K (ND280), NO$\nu$A, MINER$\nu$A, $\mu$BooNE, and DUNE experiments. Using pure calorimetric reconstruction, our results show significant differences in the mean neutrino-energy reconstruction bias between models, at the level of $\sim$7-9\,MeV for scintillator detectors and $\sim$11-18\,MeV for argon-based detectors in the relevant energy range. The latter is shown to be reduced (down to $\sim$3.5\,MeV) when using an idealised hybrid energy reconstruction based on tracking and calorimetry. Overall, we conclude that neutrino-energy reconstruction bias due to material effects may imply non-negligible systematic uncertainties for neutrino oscillation and cross-section measurements, and discuss alternative analysis strategies to mitigate the issue. 

\end{abstract}

\maketitle

\section{Introduction} 

The next generation of long-baseline neutrino oscillation experiments, Hyper-Kamiokande (Hyper-K)~\cite{Hyper-Kamiokande:2018ofw} and DUNE~\cite{abi2020longbaseline}, are poised to precisely characterise neutrino oscillations~\cite{deBlas:2025gyz} using intense beams of neutrinos and anti-neutrinos,
offering more than an order of magnitude greater statistics than the currently running experiments, T2K~\cite{abe2011t2k} and NO$\nu$A~\cite{ayres2007nova}.
Prospects for neutrino oscillation measurements at DUNE and Hyper-K include precision constraints on the PMNS oscillation parameters~\cite{10.1143/PTP.28.870,Bilenky:1978nj,ParticleDataGroup:2024cfk}, a potential discovery of $CP$-violation in the leptonic sector, and searches for new physics beyond the Standard Model~\cite{deBlas:2025gyz}. 
Since oscillation probabilities evolve characteristically in neutrino energy, its accurate reconstruction from final-state products of neutrino interactions with the nuclei within detectors is of fundamental importance for experimental success. 

Experiments employ a variety of techniques for neutrino energy reconstruction, motivated by differences in detector technology and beam energy~\cite{NuSTEC:2017hzk,dolan2026cpviolation,dolan2026characterising}: T2K~\cite{abe2011t2k} and the upcoming Hyper-K experiment~\cite{Hyper-Kamiokande:2018ofw} employ a Water-Cherenkov tank as a Far Detector (FD) to study a neutrino beam peaked at 0.6\,GeV. The magnetised scintillator-tracker ND280 acts as a Near Detector (ND), characterising the flux and neutrino interactions prior to oscillations. Its recent upgrade added the 3D-segmented SuperFGD, extending sensitivity to hadron kinematics in neutrino final states~\cite{T2K:2026zms}.
It will be accompanied by a Cherenkov-based Intermediate Detector (IWCD)~\cite{Hyper-Kamiokande:2025asb} in the Hyper-K era. The NO$\nu$A experiment~\cite{ayres2007nova} instead deploys a mineral-oil-based liquid scintillator detector at both ND and FD, which are exposed to a neutrino beam peaked at 1.9\,GeV. The MINER$\nu$A experiment measured cross sections with a scintillator-based tracker in the NuMI beam, with fluxes peaked around 3.5\,GeV in the low-energy configuration (NuMI-LE) and 6\,GeV in the medium-energy configuration (NuMI-ME)~\cite{aliaga2014design}. $\mu$BooNE~\cite{acciarri2017design}, SBND~\cite{SBND:2025lha}, and ICARUS~\cite{ICARUS:2023gpo} employ Liquid Argon Time Projection Chambers (LArTPCs) in the Booster Neutrino Beam, peaked at 0.8\,GeV, as will the upcoming DUNE experiment~\cite{abi2020deep, abi2020longbaseline} which will use the LBNF beam peaked at 2.2\,GeV.

T2K and Hyper-K apply a neutrino energy reconstruction approach based only on kinematics of charged
leptons for neutrino oscillation measurements, motivated by the characteristics of neutrino interactions at the comparatively low beam energy, and the
technology of the FD in which particles below the Cherenkov threshold are invisible~\cite{abe2011t2k, Hyper-Kamiokande:2018ofw}.
In contrast, the MINER$\nu$A, NO$\nu$A,
$\mu$BooNE and DUNE experiments adopt a different approach, utilising high granularity tracking detectors to reconstruct a detailed image of the final state of the neutrino interaction~\cite{ayres2007nova,
adamson2017measurement, aliaga2014design, acciarri2017design,
abi2020longbaseline, abi2020deep}. Although some hadrons are not reconstructable as clean tracks, they still leave observable energy deposits localised around the neutrino interaction vertex (commonly referred to as \textit{vertex activity}).
The neutrino energy is then effectively reconstructed as the sum of energies of observed particle tracks and vertex activity, thereby applying a calorimetric approach in which the visible charge observed in the detector is related to the energy deposited. 

Relating reconstructed to true neutrino energy requires corrections in both the calorimetric and kinematic methods, which are built from theoretical models describing neutrino interactions with nuclei~\cite{dolan2026characterising,NuSTEC:2017hzk}.
Model predictions of this relation vary greatly in the phase space of interest to neutrino oscillation experiments, rendering potential mis-modelling of neutrino interactions a dominant projected contribution to systematic uncertainties at DUNE and Hyper-K~\cite{dolan2026cpviolation}. Recent work suggests that the mapping between true and reconstructed neutrino energy must be constrained at the $\sim$0.5\%-level to fully reach the precision goals of next-generation experiments~\cite{dolan2026characterising}.

Motivated partially by this challenge, a detailed campaign of neutrino interaction measurements is underway. Many of these also use calorimetric approaches to reconstruct neutrino energy or the total hadronic energy, for example in recent results published by MINER$\nu$A and $\mu$BooNE~\cite{ascencio2022measurement, ruterbories2022simultaneous, abratenko2024measurement}. In these measurements, the region of low energy transfer, $q_0$, was found to be particularly prone to modelling problems, a region in which the untracked vertex activity tends to make up a significant part of the hadronic energy, as comparatively less energy is available to hadrons. 

One important but rarely-discussed effect in the context of using calorimetric approaches in oscillation or cross-section measurements the is the non-linear correspondence
between energy loss and visible energy in scintillators and LAr-TPCs due to quenching and recombination effects, respectively. 
These effects are particularly pronounced at high stopping
powers, that is, in Bragg peaks of stopping particles, and thus strongly
affect vertex activity where the relative contribution of energy deposits coming from Bragg peaks is largest. Since predictions of hadron kinematics differ between
interaction models, calorimetrically reconstructed quantities are inherently
biased towards the model used to build the relation between quenched charges
and the sum of contributing visible particle energies.

The characterisation of model-dependencies introduced when correcting for material effects in calorimetric variables
represents the aim of this paper.  Predictions from different
state-of-the-art interaction models are prepared for fluxes and
detector materials relevant to the T2K/Hyper-K~\cite{abe2011t2k,Hyper-Kamiokande:2018ofw}, NO$\nu$A~\cite{ayres2007nova}, MINER$\nu$A~\cite{aliaga2014design}, $\mu$BooNE~\cite{acciarri2017design}, and DUNE experiments~\cite{abi2020deep}. Section~\ref{sec:generators} provides a summary of the interaction models used in this comparison.
A simulation of the relevant material effects is applied as described in Section~\ref{sec:material_effects}, and a calorimetric energy reconstruction is
prepared for a reference model and then applied on the set of alternative models, which is detailed in Section~\ref{sec:calo_vars}. We compare a pure calorimetric reconstruction approach; one based on particle tracking alone, assuming perfect reconstruction above tracking thresholds; and a hybrid approach where tracking is combined with a calorimetric reconstruction of untracked particles. The
model-dependent bias due to the aforementioned material effects is
evaluated by comparing reconstructed quantities to results obtained in the absence of such effects, and results are shown in Section~\ref{sec:results} and discussed in Section~\ref{sec:discussion}. Finally, we discuss alternative forward-folding-based alternative analysis approaches in Section~\ref{sec:alt_appr}.

\section{\label{sec:generators}Neutrino Interaction Models} 

Charged-Current interactions of muon neutrinos ($\nu_\mu$CC) are simulated at the upgraded T2K/Hyper-K Near Detector ND280, MINER$\nu$A, $\mu$BooNE, as well as the NO$\nu$A and DUNE Far Detectors. For ND280 and MINER$\nu$A we consider interactions in polystyrene-based scintillators (C$_8$H$_8$), while interactions on Mineral-oil-based liquid scintillator (CH$_2$) are simulated for NO$\nu$A. For the LAr-TPCs $\mu$BooNE and the DUNE FD interactions are simulated on Argon. 

For the two FDs, NO$\nu$A and DUNE, we consider oscillated $\nu_\mu$ fluxes~\cite{Adamson:2015dkw,abi2020deep}, using oscillation parameters from the NuFit 5.2 best fit values~\cite{Esteban:2020cvm,nufitweb}. For the T2K experiment, calorimetric reconstruction approaches are only possible at ND280, motivating the choice of the ND flux~\cite{T2K:2012bge}. For MINER$\nu$A we chose the NuMI-LE flux~\cite{Adamson:2015dkw}, and for $\mu$BooNE the BNB flux~\cite{MicroBooNE:2023foc,MiniBooNE:2008hfu}, the latter being limited to a maximal neutrino energy of 3\,GeV in this paper.

This work uses the full set of interaction models detailed in Ref.~\cite{dolan2026cpviolation}, including different tunes of
the GENIE~\cite{andreopoulos2009genie,
andreopoulos2015geniemanual, alvarez2021recent}, GiBUU~\cite{buss2012transport}, NuWro~\cite{golan2012nuwro,golan2012effects,niewczas2019nuclear}, as well as NEUT~\cite{hayato2001neut,hayato2009neutrino,hayato2021neut} generators, via NUISANCE~\cite{Stowell:2016jfr}.
They are chosen as a representative sample of current state-of-the-art and widely used neutrino interaction generators and such that their spread represents a broad envelope for the scale of plausible variation of neutrino interaction modelling.
The use of model spread, rather than considering some variation of uncertainty parameters within a single model, is motivated by the observation that no single model consistently describes experimental data, nor does any provide a complete and well-defined uncertainty prescription that fully captures all of the relevant nuclear physics~\cite{Avanzini:2021qlx, dolan2026cpviolation}. 

In addition to the models considered in Ref.~\cite{dolan2026cpviolation}, we consider one additional generator configuration labelled as GENIE~\texttt{G18\_10d}. 
This is equivalent to any of the \texttt{G18\_10a/b/c} configurations but with an alternative approach to describing final-state interactions (FSI) using the GEANT4 Bertini Cascade model (G4BC)~\cite{Heikkinen:2003sc}. 
This configuration is added as it uses a sophisticated FSI model to predict significantly different hadronic final states to other model configurations, which include contributions from nuclear clusters (e.g. $\alpha$ particles) that can significantly impact biases caused by non-linear material effects.

The envelope spanned by this set of models therefore offers a pragmatic approach to account for the variation that our incomplete knowledge of neutrino interactions could introduce. 
However, it should be stressed that this model spread is not a substitute for a formal uncertainty and, given that many models share ingredients or assumptions, should likely be considered as a only lower bound on plausible variation.

A summary including references to all contributions of the interaction models compared here is provided in Tables~\ref{tab:models} and~\ref{tab:fsi} in the Appendix.

\begin{figure}[t]
    \includegraphics[width=.4\textwidth]{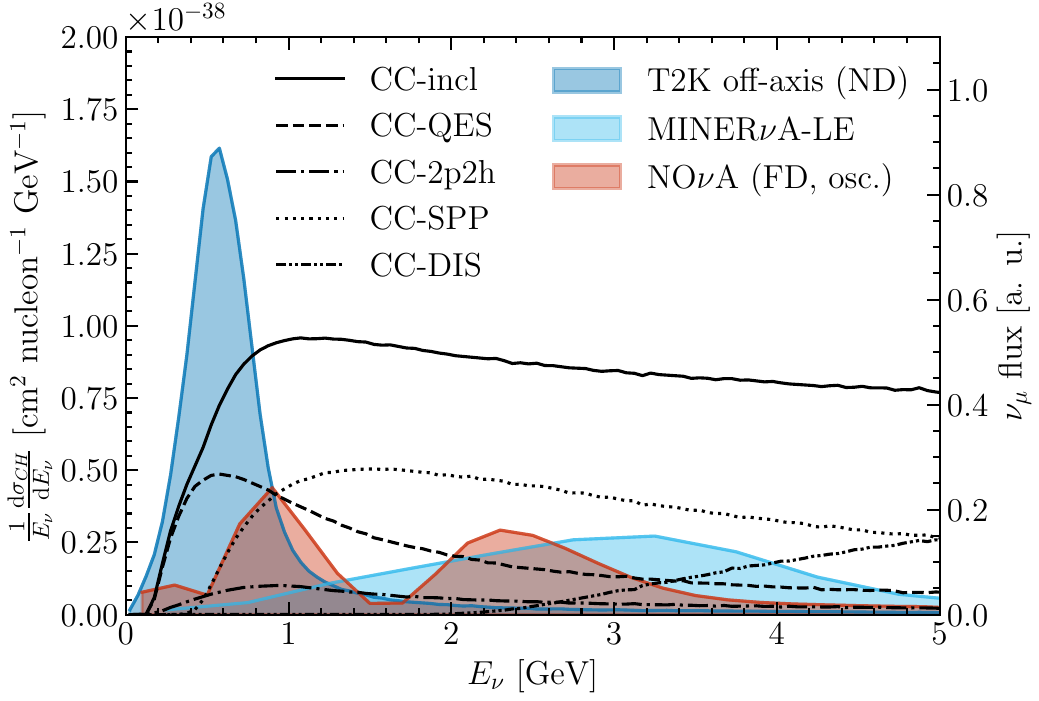}
    \includegraphics[width=.4\textwidth]{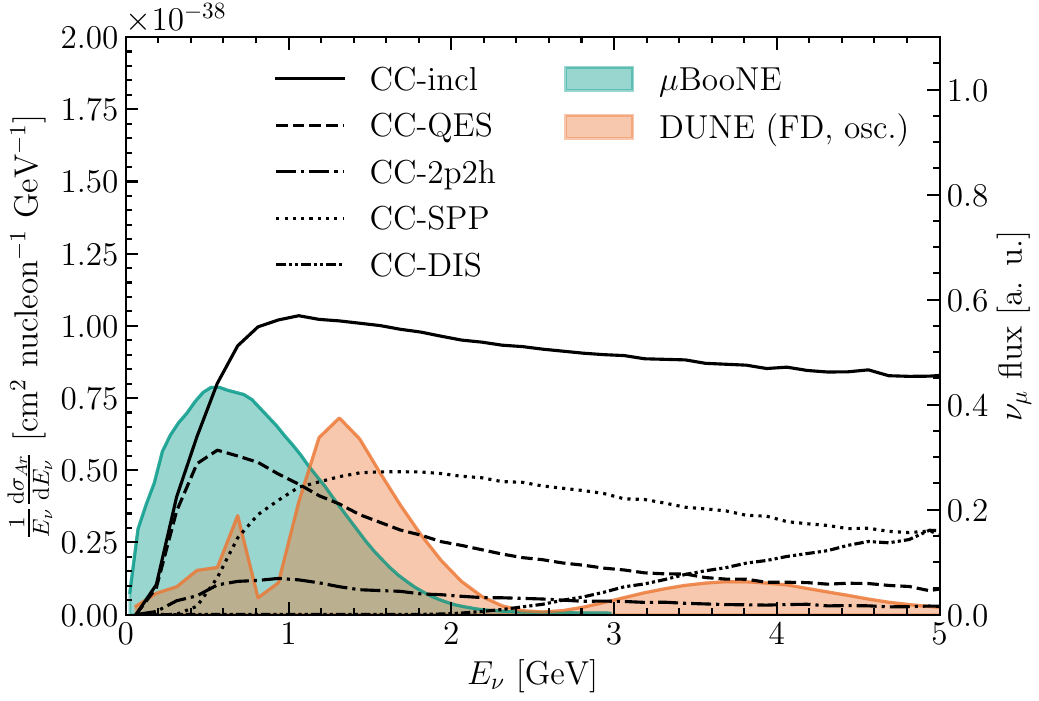}
    \caption{\label{fig:flux} Interaction cross-sections on Hydrocarbon (top) and Argon (bottom) as predicted by GENIE
    \texttt{G18\_10a}, as well as $\nu_\mu$ fluxes at the
    T2K ND, MINER$\nu$A, $\mu$BooNE, and the NO$\nu$A and DUNE FD (oscillated).}
\end{figure}

Charged-current neutrino interactions with nuclei receive contributions from different interaction channels~\cite{NuSTEC:2017hzk,dolan2026characterising}. Quasi-Elastic Scattering (CC-QES) covers interactions with a single neutron that is transformed into a proton, Single Pion Production (CC-SPP) covers interactions in which a single pion is produced, primarily via the promotion of a nucleon to a resonance state which then emits a pion during its subsequent decay. Deep Inelastic Scatting (CC-DIS) covers interactions in which the energy transfer of the interacting neutrino is large enough to resolve the parton structure within nucleons, creating a hadronic shower. Finally, Two-Particle Two-Hole interactions (CC-2p2h) describe interactions with collective states of two nucleons, usually bound by meson exchange currents. The relative contribution of different interaction
modes differs between each experiment, and Figure~\ref{fig:flux} shows cross-sections as predicted by the reference
model GENIE~\texttt{G18\_10a} for relevant processes on Argon and Hydrocarbon, alongside the flux spectra used with these experiments in the following. 

\begin{figure*}[t]
    \begin{subfigure}[t]{0.48\textwidth}
    \includegraphics[height=.60\textwidth]{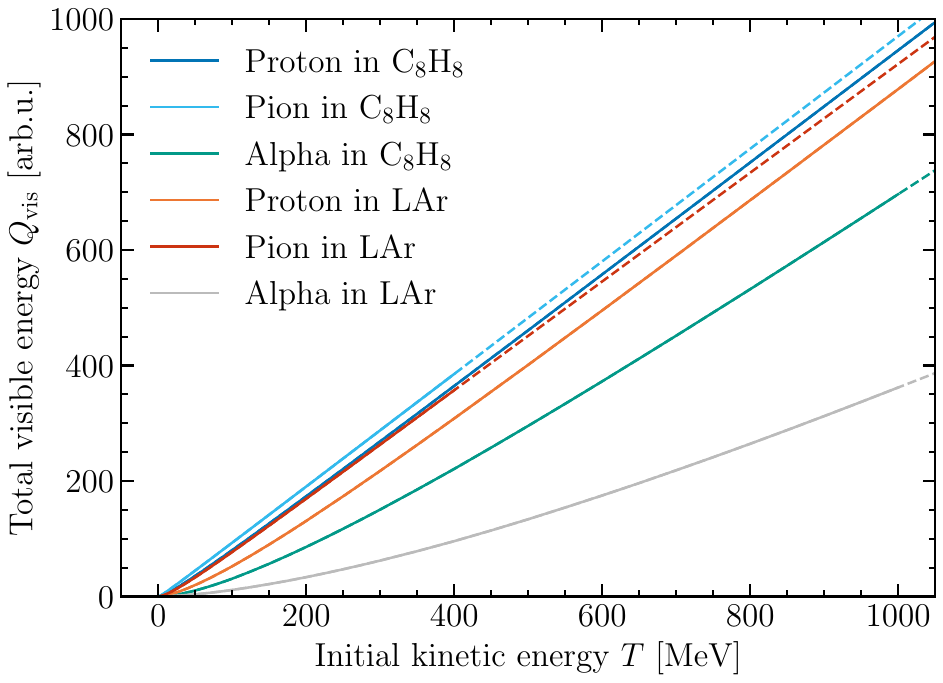}
    \caption{ }
    \end{subfigure}
    \begin{subfigure}[t]{0.48\textwidth}
    \includegraphics[height=.60\textwidth]{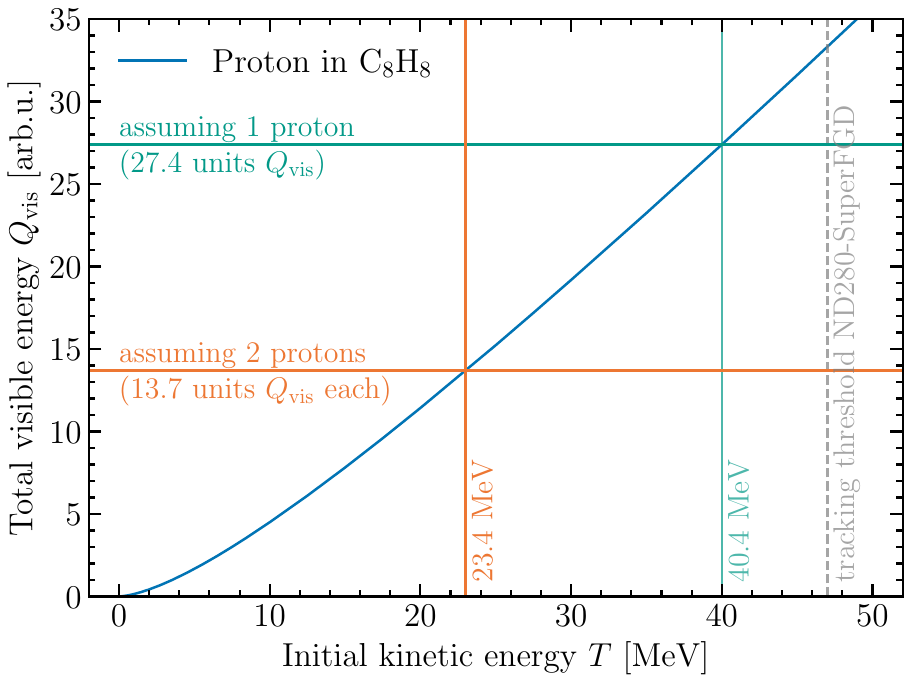}
    \caption{ }
    \end{subfigure}
    \vspace{-1em}
    \caption{\label{fig:example_1vs2p}
        (a) Integrated visible charge caused by single particles in polystyrene (C$_8$H$_8$)
        and LAr using material constants listed in
        Table~\ref{tab:materialconsts}, where $\varepsilon_\mathrm{LAr}=0.5$\,kV/cm.
        Units for visible charges are arbitrary, and to aid comparison with
        scintillators this plot uses $c_\mathrm{MIP}^\mathrm{LAr}=1.38$ while kept at unity for both materials in the simulation. Dashed
        lines indicate linear interpolation outside the range of available
        $\diff E/\diff x$ literature tables.  
        (b) Illustration of the bias introduced when interpreting a fixed
        amount of light yield as one proton of 40\,MeV (green) or as split
        between two protons of equal energy (orange), resulting in 23\,MeV per
        proton, or $\Sigma T_p$ of 46\,MeV. The gray dashed line indicates the tracking threshold of the SuperFGD at the T2K ND280.
    }
\end{figure*}

\section{\label{sec:material_effects}Non-linear material effects}

The \textit{visible charge}, $Q_{vis}$, measured in the detector forms the
basis for any calorimetric energy reconstruction.  This is typically the measured
scintillation light in the case of plastic scintillators and the collected ionisation charge in the case of TPCs. The visible charge is used as a proxy for the \textit{visible
energy}, $E_{vis}$, defined as the sum of energies across the collection of
particles that left visible traces in the detector, thereby causing the given amount of visible charge.  The correspondence
between the visible charge and visible energy is not linear at high stopping powers, as energy deposits in
plastic scintillators are subject to quenching~\cite{birks1951scintillations},
and ionisation charge recombination effects come into play in
TPCs~\cite{acciarri2013study}.  

\subsection{Empirical parametrisation}

Birks' law is commonly used to describe the
functional dependence of visible scintillation light, $Q_\mathrm{scint}$, on
the energy loss of the particle, $E$, in a small track segment $\diff
x$~\cite{birks1951scintillations}:

\begin{equation}
    \label{eq:birkslaw} 
    \frac{\diff Q_\mathrm{scint}}{\diff x} =
    c_\mathrm{MIP}\cdot\frac{1}{1+c_\mathrm{B}\frac{\diff E}{\diff x}} 
    \cdot \frac{\diff E}{\diff x}.
\end{equation} 

The material constant $c_\mathrm{B}$ is referred to as \textit{Birks constant},
and a general detector response factor $c_\mathrm{MIP}$ is included in this
equation to quantify the linear part of the detector response, typically inferred from
well-known $\diff E/\diff x$ values of Minimally Ionising Particles (MIPs). 

In the case of TPCs, Birks law can also be used to approximate the relationship between visible charge and energy loss
at first order. However,
most experiments use a more precise alternative, commonly referred to as
the \textit{modified Box law}~\cite{acciarri2013study}: 

\begin{equation} 
    \label{eq:boxlaw} 
    \frac{\diff Q_\mathrm{ion}}{\diff x} =
    c_\mathrm{MIP}\cdot\frac{1}{b_\mathrm{box}\frac{\diff E}{\diff x}} \cdot
    \ln\left(a_\mathrm{box} + b_\mathrm{box}\frac{\diff E}{\diff x} \right) 
    \cdot \frac{\diff E}{\diff x}.
\end{equation}

Here, empirical values for the factors ${a_\mathrm{box}=0.930}$ and
${b_\mathrm{box}=0.212/(\varepsilon\rho)}$, taken from \citet{acciarri2013study},
maintain a dependence on the product of the electric field, $\varepsilon$, and the
density of the material, $\rho$.  

In both cases, this non-linear correspondence leads to ambiguities when an unknown amount of particles
contribute to the detected visible charge $\diff Q$ in a small
segment $\diff x$. While energy deposits occupying the same microscopic volume in general have a negligible impact on the amount of quenching or recombination in the detectors
discussed here, biases can be introduced when relating the visible charge to a sum of particle energies if particle multiplicities and relative energy contributions
are misjudged or can not be resolved, such as in vertex activity. Assumptions on particle kinematics and multiplicities are required to infer visible energies from visible charges, introducing a model-dependent bias.

Figure~\ref{fig:example_1vs2p}~(a) shows the total visible charge caused by various particles as a
function of its initial kinetic energy in scintillators
(polystyrene) and LAr-TPCs (more details of the simulation used for these figures follow
in the latter half of this section). The non-linear behaviour is visible at
energies close to the tracking thresholds of these detectors at few tens of MeV.
Figure~\ref{fig:example_1vs2p}~(b) shows an example scenario where a fixed
amount of light yield in a scintillator is interpreted as one proton, or as two
protons which each contribute half of the light yield. The corresponding sums of proton kinetic energies
are 40\,MeV and 46\,MeV, respectively. This ambiguity gives rise to a bias which depends on highly interaction-model-dependent multiplicity assumptions, and the quantification of this bias is the aim of this paper. The situation is complicated further as not only protons, but also pions and, as predicted by some models, heavy nuclear clusters contribute to the hadronic energy in neutrino interactions.

\subsection{\label{sec:sim}Simulation} 

For the purpose of the study presented in this paper, a simplified detector
response simulation was prepared that only includes Birks quenching in scintillators and charge recombination in LAr-TPCs, while ignoring all other effects that have to be accounted
for in a realistic detector simulation.  Birks' Law (Eq.~\ref{eq:birkslaw})
is used to simulate the correspondence between energy loss and resulting light
yield for scintillator-based detectors (T2K's upgraded ND280, the NO$\nu$A FD,
and MINER$\nu$A), and the modified Box law (Eq.~\ref{eq:boxlaw}) is used for
ionisation charges in LAr-TPCs ($\mu$BooNE and the DUNE FD, in particular the horizontal drift module at the target electric field value of 0.5\,kV).  Corresponding material parameters are summarised in Table~\ref{tab:materialconsts}.

The total visible charge $Q_\mathrm{vis}$ deposited by a particle of given initial
kinetic energy can be calculated as the integral of visible charges $\diff Q/\diff x$ caused by energy losses $\diff Q/\diff x$ in small steps along the track, until
the particle comes to a stop:

\begin{equation}
\label{eq:integr_charge}
    Q = \int \frac{\diff Q}{\diff x}\left( \frac{\diff E}{\diff x}\right) \diff x,
\end{equation}

where the expected functional dependence of  $\diff Q/\diff x$ on $\diff E/\diff x$ is given by either Birks law (Eq.~\ref{eq:birkslaw}) for scintillators or Box law (Eq.~\ref{eq:boxlaw}) for LAr-TPCs.

To determine the total visible charge deposited by a particle of given initial kinetic energy, a numerical integration of Equation~\ref{eq:integr_charge} is applied, using literature
values for stopping powers $\diff E/\diff x$ published by the National Institute of Standards and
Technology (NIST)~\cite{berger1993stopping, berger1999star}. For simplicity, only
protons, charged pions, and nuclear clusters are subjected to material effects,
while energies of all other particles (primary lepton, neutral pions and other
mesons, photons, and rarer particles) are kept unmodified throughout simulation and
reconstruction. Material effects for nuclear clusters heavier
than deuterons, that is, with a nuclear charge of $Z>1$, are treated as
alpha particles of the given kinetic energy, and deuterons are treated as
protons. The corresponding  $\diff E/\diff x$ is thus underestimated in these cases~\footnote{This can be
inferred from the charge dependence of the Bethe-Bloch equation~\cite{ParticleDataGroup:2024cfk}.}, leading to an underestimation of the amount of quenching or recombination effects for such particles. Results for models that predict a liberation of such heavy nuclear clusters are thus conservative estimates. It should be further noted that all particles are assumed to stop within the detector volume, meaning that no pion-decays in flight are accounted for. This would further reduce the contribution of visible charge from pions compared to that of protons or nuclear clusters.

\begin{table}[t]
    \centering
    \renewcommand*{\arraystretch}{1.1}
    \begin{tabular}{l c c c}
        \hline  \hline
         & Constant & Value & Ref. \\\hline
         \multirow{2}{*}{\rotatebox[origin=c]{90}{C$_8$H$_8$}} 
                & $c_B$ & 0.0126\,cm/MeV & \cite{leverington2011scintillating} \\
                & $\rho$& 1.060\,g/cm$^3$ & \cite{pdg2024polystyrene}\\\hline
        \multirow{2}{*}{\rotatebox[origin=c]{90}{CH$_2$}} 
                & $c_B$ & 0.0120\,cm/MeV & \cite{acero2023measurement} \\
                & $\rho$& 0.862\,g/cm$^3$ & \cite{mufson2015liquid}\\\hline
         \multirow{5}{*}{\rotatebox[origin=c]{90}{LAr}} 
                & $a_{box} $& 0.93 & \cite{acciarri2013study}\\
                & $b_{box}$& 0.307 cm/MeV & \cite{acciarri2013study,bnl_lar_notes}\\
                & $\rho$& 1.38\,g/cm$^3$ & \cite{acciarri2013study} \\
                & $\varepsilon_{\mu\mathrm{BooNE}}$& 0.273\,kV/cm &
                \cite{acciarri2017noise}\\
                & $\varepsilon_\mathrm{DUNE}$& 0.5\,kV/cm & \cite{abi2020deep_iv}\\
         \hline\hline
    \end{tabular}
    \caption{\label{tab:materialconsts} Literature values for material constants describing typical
        polystyrene-based scintillators such as used by T2K and MINER$\nu$A (C$_8$H$_8$), mineral-oil-based scintillators such as that of NO$\nu$A (CH$_2$), and liquid Argon (LAr) used in TPCs. The overall scaling constant $c_\mathrm{MIP}$ is
        set to 1 for both materials throughout this study.
        }
\end{table}

\section{\label{sec:calo_vars}Calorimetric energy reconstruction} 
 
We consider three variants of calorimetric reconstruction approaches, and compare them to a pure tracking-based approach: 
firstly, pure calorimetry that does not rely on particle identification or energy reconstruction of individual hadrons; secondly a similar approach that applies this method separately on observable topologies (split by presence of at least one charged pion), and a third approach that assumes perfect energy reconstruction of individual particles above the tracking threshold of a detector, while applying the calorimetric method to untracked particles only. 

This section provides detailed variable definitions used in the following bias evaluation, as well as a description of the approaches applied in order to reconstruct them. All calorimetric results are based on GENIE \texttt{G18\_10a} as the reference model used to build the reconstruction.

\subsection{Variable definitions} 

The \textit{visible hadronic energy}, $E_\mathrm{had,vis}$, is defined as the sum of energies
of all particles that cause visible traces in the  scintillator or LAr-TPC. For neutrino interactions, this typically contains the following contributions:

\begin{equation}
    \label{eq:ehadvis_true}
    E_\mathrm{had,vis} = \Sigma T_{p}+\Sigma T_{\pi^\pm}+\Sigma T_\alpha +
    \Sigma E_{\pi^0,\gamma,\mathrm{other}},
\end{equation}

where $T_\alpha$ stands representative for all nuclear clusters.
Protons and nuclear clusters ($\Sigma T_p$ and $\Sigma T_\alpha$, respectively) contribute as
kinetic rather than total energies since nucleons are already present in the
nucleus, meaning that the mass already existed prior to the interaction, and since they do not undergo a decay following their
liberation (at relevant timescales). Moreover, kinetic energies are
usually also chosen for charged pions ($\Sigma T_{\pi^\pm}$) in the definition of this observable, since part of the energy goes unseen in their decay. Neutral pions, photons, and all other, rare types of particles created in neutrino interactions are added at their total energy value ($\Sigma E_{\pi^0,\gamma,\mathrm{other}}$).

If $E_\mathrm{had,vis}$ were to be interpreted as a proxy for the energy
transfer $q_0$, it should be noted that alongside missing pion masses also
kinetic energies of neutrons and other invisible neutral particles created in the interaction are not
accounted for.  A truth-level quantity that includes these contributions is
often constructed as the (total) hadronic energy $E_\mathrm{had}$.  A
comparison of $E_\mathrm{had,vis}$ and $E_\mathrm{had}$ to $q_0$ can be found in Refs.~\cite{wilkinson2022substandard} and~\cite{dolan2026cpviolation}. The study presented here instead aims to
evaluate the intrinsic bias in $E_\mathrm{had,vis}$ introduced during the
reconstruction step when visible charges $Q_\mathrm{had,vis}$ are interpreted
as sums of corresponding charged particle energies.  All bias estimates shown in the
following are evaluated against the true value of $E_\mathrm{had,vis}$ and are
thus by construction not affected by biases due to missing neutron
energies or pion masses.

\begin{table}[t]
    \centering
    \renewcommand*{\arraystretch}{1.3}
    \begin{tabular}{l c c r}
        \hline\hline
        Experiment & $T_p^\mathrm{min}$ [MeV]& $T_{\pi^\pm}^\mathrm{min}$ [MeV] & Ref.\\\hline
        T2K (SuperFGD)& 47 & 30 & \cite{T2K:2026zms,nd280up_tdr}\\
        MINER$\nu$A & 100 & 35 & \cite{Avanzini:2021qlx} \\
        $\mu$BooNE & 47 & 30 & \cite{abratenko2020measurement,Avanzini:2021qlx,microboone2017pandora}\\
        DUNE (FD) & 40 & 30 & \cite{abi2020deep_iv}\\
    \hline\hline
    \end{tabular} \caption{\label{tab:tracking_thresholds}Tracking thresholds
    for proton and pion kinetic energies, and corresponding references.}
\end{table}

Highly segmented scintillators and LAr-TPC-based neutrino detectors offer unprecedented tracking capabilities and allow to resolve most contributions to $E_\mathrm{had,vis}$  as individual particle tracks~\cite{T2K:2026zms,nd280up_tdr,Avanzini:2021qlx,abratenko2020measurement,microboone2017pandora,abi2020deep_iv}. Table~\ref{tab:tracking_thresholds} provides values of tracking thresholds for protons and charged pions in detectors that apply particle tracking as part of their analyses. The sum of visible hadronic energies of particles above the tracking threshold is defined as $E_\mathrm{had,vis}^{tracked}$, and the remainder of untracked visible hadronic energy contributions in vertex activity as $E_\mathrm{had,vis}^\mathrm{VA}$:

\begin{equation}
    E_\mathrm{had,vis}=E_\mathrm{had,vis}^{tracked}+E_\mathrm{had,vis}^\mathrm{VA}.
\end{equation}

In general, we denote contributions of particles below the tracking threshold of a given detector with the index \textit{VA}.

Distributions of true $E_\mathrm{had,vis}$ for the set of interaction models considered in this work and the model spread in the relative content of vertex activity ($E_\mathrm{had,vis}^\mathrm{VA}/E_\mathrm{had,vis}$) can be found in the Appendix (Figures~\ref{fig:ehadvis_cc0pi} and~\ref{fig:ehadvis_cc1pi}).

\subsection{Inference of visible energies from visible charges} 

\begin{figure*}[t]
  \begin{center}
  \hspace{0.1\textwidth}%
  \begin{subfigure}{0.33\textwidth}\includegraphics[width=\textwidth,trim={0 0 0 5.5mm},clip]{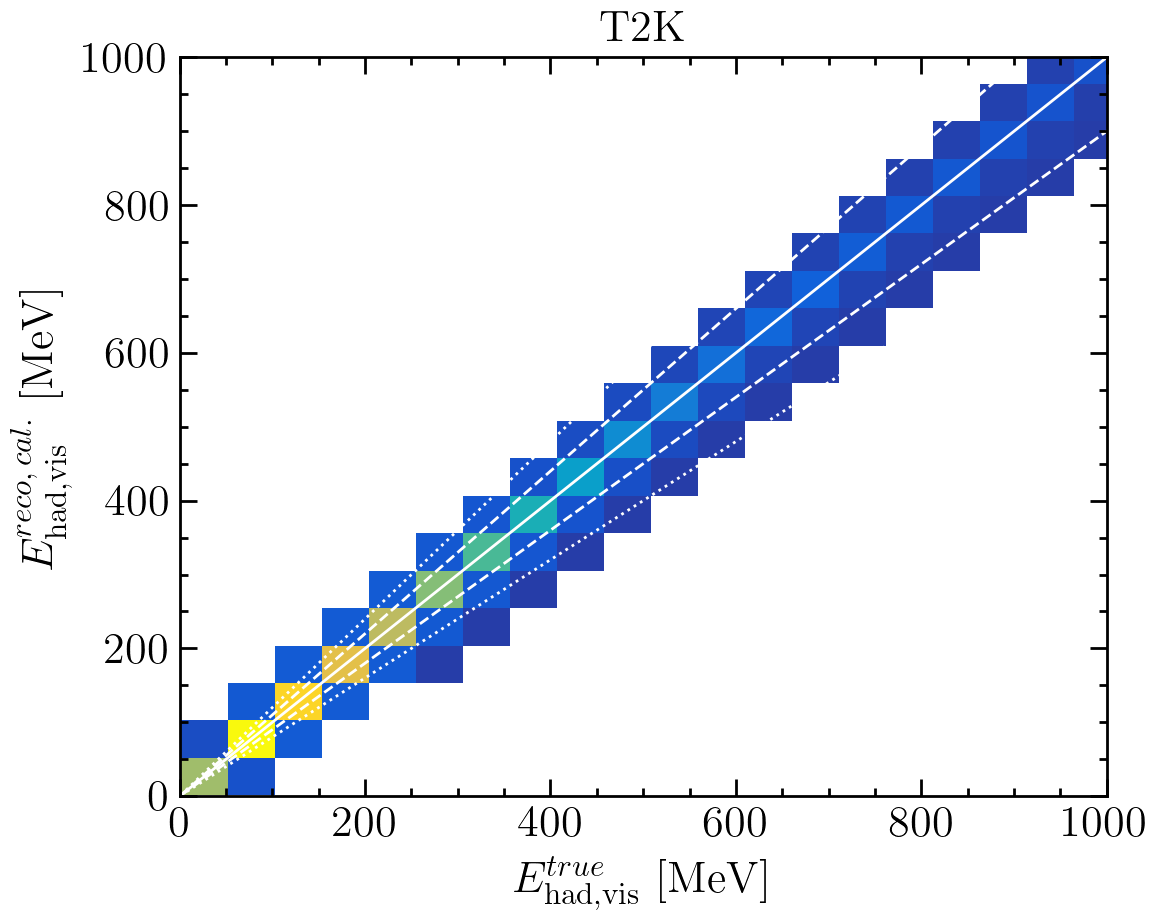}
  \caption{T2K-ND280}
  \end{subfigure}%
  \begin{subfigure}{0.33\textwidth}\includegraphics[width=\textwidth,trim={0 0 0 5.5mm},clip]{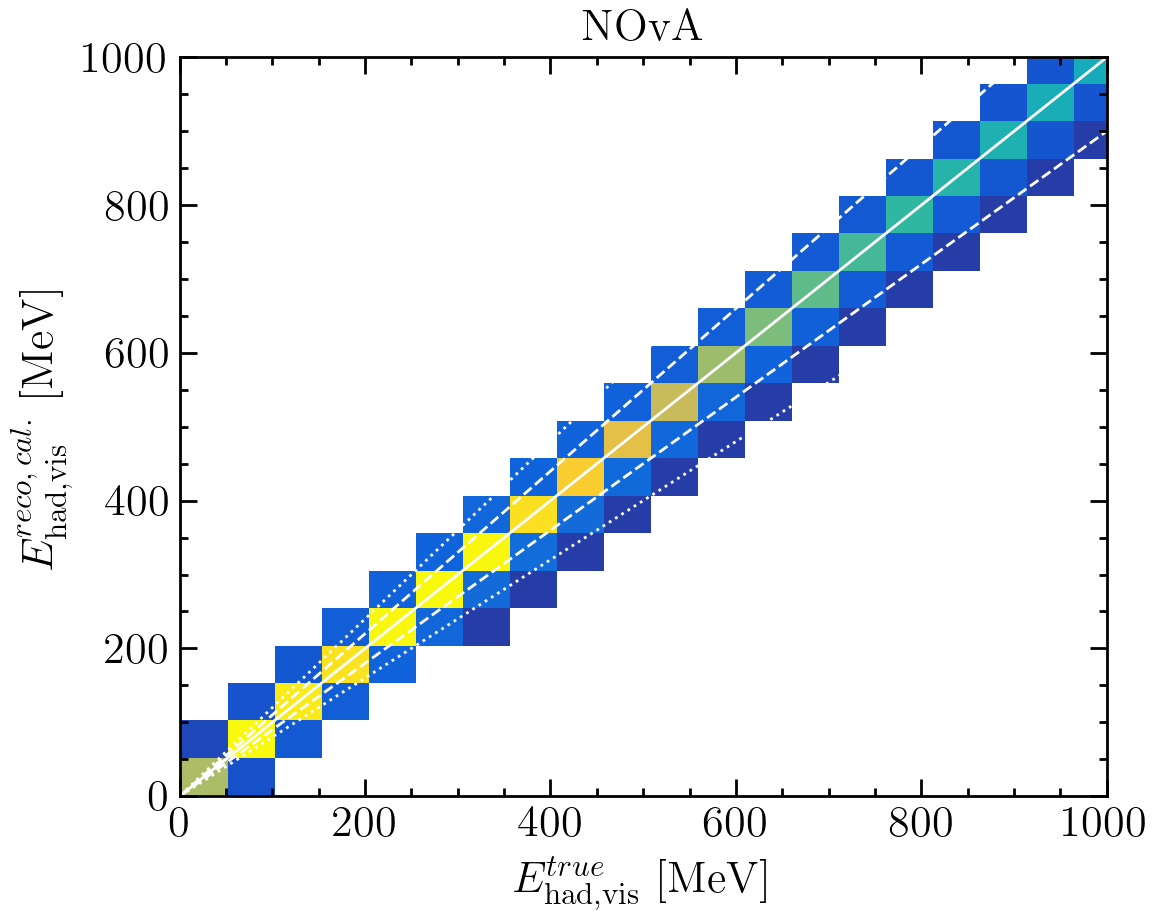}
  \caption{NO$\nu$A}
  \end{subfigure}%
  \begin{subfigure}{0.33\textwidth}\includegraphics[width=\textwidth,trim={0 0 0 5.5mm},clip]{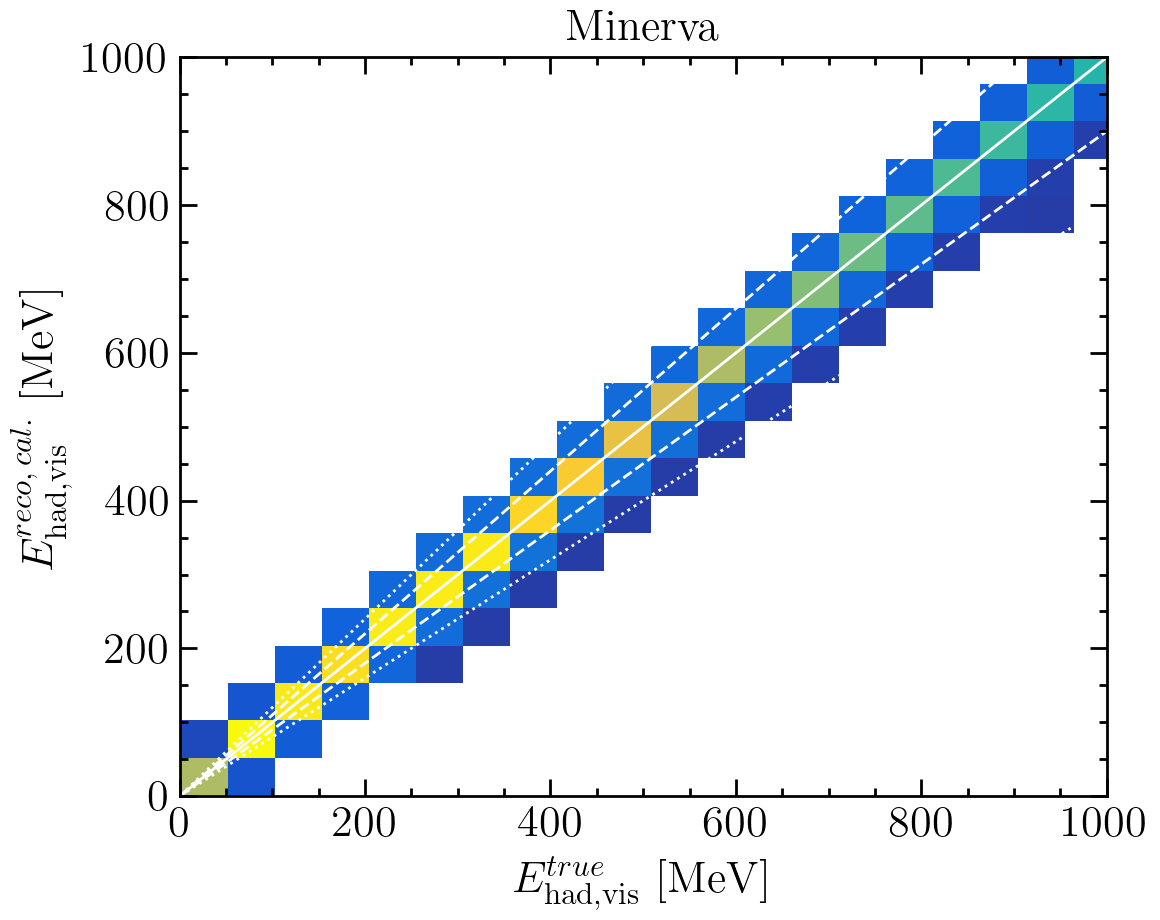}
  \caption{MINER$\nu$A}
  \end{subfigure}
  
  \vspace{0.5em}
  
  \begin{subfigure}{0.33\textwidth}\includegraphics[width=\textwidth,trim={0 0 0 5.5mm},clip]{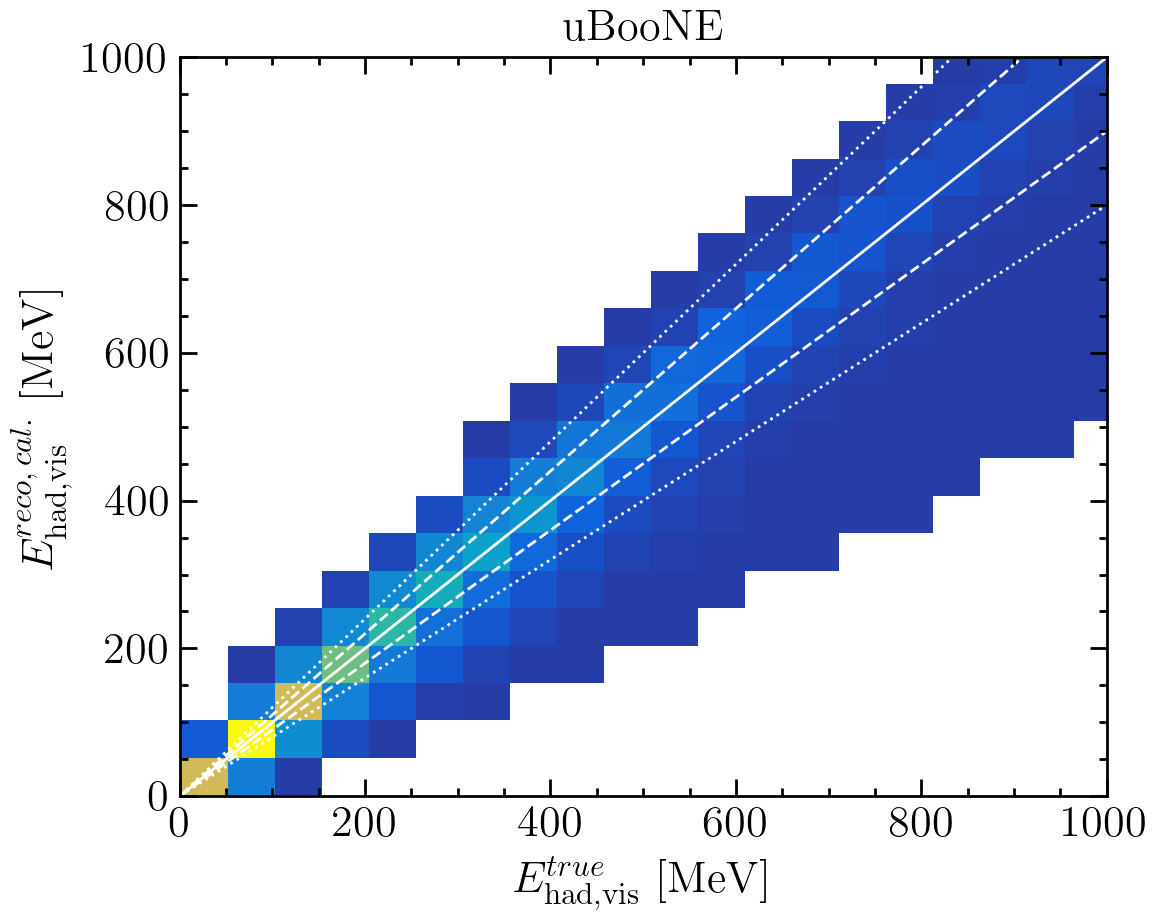}
  \caption{$\mu$BooNE}
  \end{subfigure}%
  \begin{subfigure}{0.33\textwidth}\includegraphics[width=\textwidth,trim={0 0 0 5.5mm},clip]{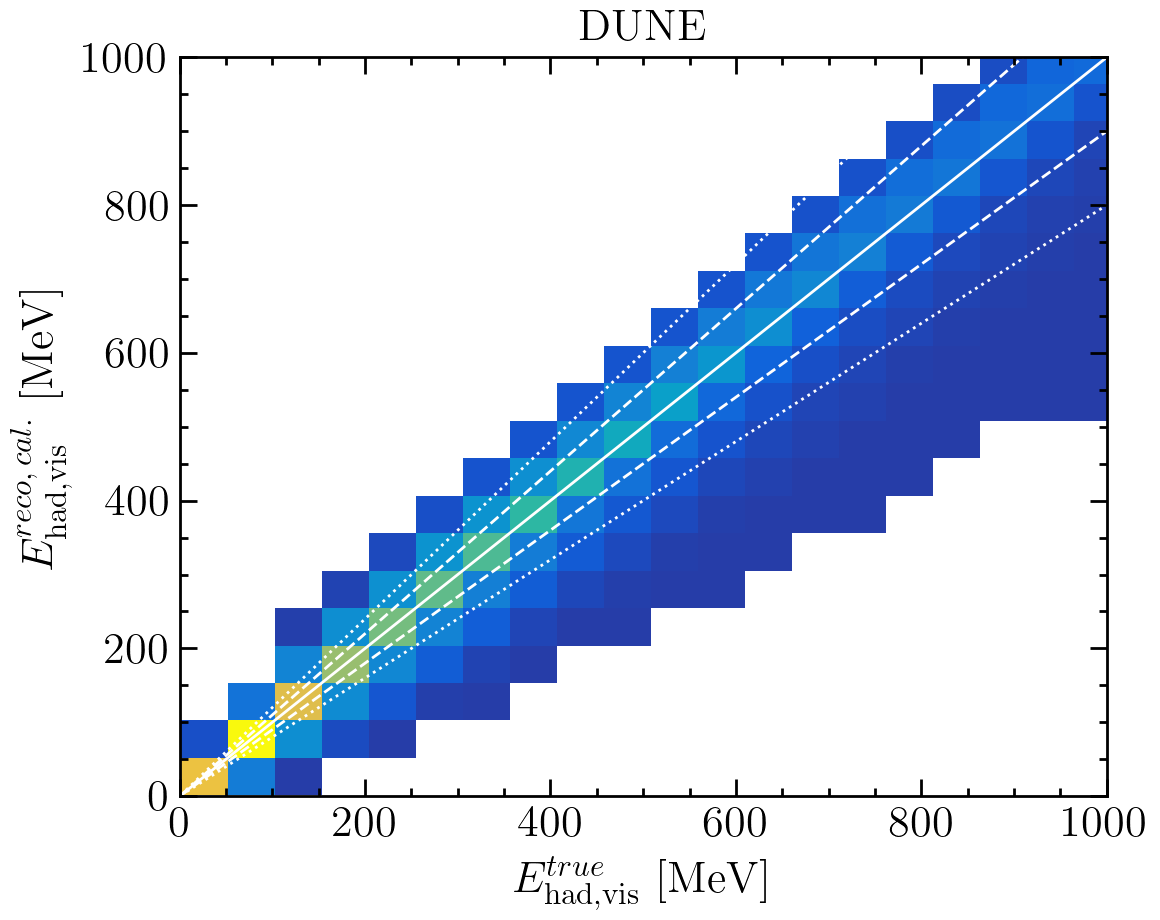}
  \caption{DUNE}
  \end{subfigure}
  \end{center}
  \vspace{-1.5em}
  \caption{\label{fig:ehadvis_smear} 
      Smearing matrices for $E_\mathrm{had,vis}^{reco,\,cal.}$ in CC-inclusive (CC-Incl) events of the
      reference generator model~(GENIE~\texttt{G18\_10a}) showing scintillator-based detectors in the top row and LAr-TPCs in the bottom row. Overlaid white dashed (dotted) lines indicate 10\% (20\%) deviation from the true value.
  }
\end{figure*}

The calorimetric reconstruction of visible hadronic energy from the measured visible charge $Q_\mathrm{had,vis}$ requires a mapping between the two quantities.
Response matrices relate the average $Q_\mathrm{had,vis}$ to the corresponding average
value of $E_\mathrm{had,vis}$, based on predictions by a
reference generator.

\subsubsection{CC-inclusive-based calorimetry}

To avoid dependence on selection performance for exclusive final states which require the identification of hadrons, such response matrices are often built for $\nu_\mu$CC-inclusive samples, i.e. requiring only the identification of the primary muon.

GENIE \texttt{G18\_10a} is used as the reference generator in this study to construct such response matrices for each experiment. The visible charge is determined on an event-by-event basis via
application of the relevant equation describing detector effects for all contributing particles, using their true kinetic energy (see response curves in Figure~\ref{fig:example_1vs2p}a) to determine the sum of all visible charges $\Sigma Q_{p,\pi^\pm,\alpha}$:

\begin{equation}
    \Sigma Q_{p,\pi^\pm,\alpha} = Q_p + Q_{\pi^\pm} + Q_\alpha.
\end{equation}

The response matrix is built to relate a given sum of hadronic visible charges $\Sigma
Q_{p,\pi^\pm,\alpha}$ to the corresponding average sum of true hadron kinetic energies, $\Sigma T_{p,\pi^\pm,\alpha}$. This correspondence can differ between experiments due to the different composition of the final state of the neutrino interaction in the CC-inclusive sample, and Figure~\ref{fig:lookup_avg_refgen} in the Appendix shows $\Sigma
Q_{p,\pi^\pm,\alpha}$ as a function of  $\Sigma
T_{p,\pi^\pm,\alpha}$ for all five experiments.

We define the total reconstructed value of $E_\mathrm{had,vis}$ via the reconstructed value of the sum of initial kinetic energies $\Sigma
T_{p,\pi^\pm,\alpha}$ given $\Sigma
Q_{p,\pi^\pm,\alpha}$,  keeping all other particle energies at truth value:

\begin{equation}
\label{eq:ehadvis_reco}
E_\mathrm{had,vis}^{reco,\,cal.} = \Sigma T^{reco}_{p,\pi^\pm,\alpha} (\Sigma Q_{p,\pi^\pm,\alpha})
    +  \Sigma E_{\pi^0,\gamma,\mathrm{other}}^{true}.
\end{equation}

Figure~\ref{fig:ehadvis_smear} shows smearing matrices of $E_\mathrm{had,vis}^{reco,\,cal.}$ against $E_\mathrm{had,vis}^{true}$ for the full CC-inclusive sample in all five experiments for the reference model. A higher off-diagonal contribution is visible for the two LAr-TPCs, $\mu$BooNE and DUNE, compared to the scintillator-based detectors.

\subsubsection{Topology-specific calorimetry}

Since FSI can alter the multiplicity, kinematics and species of particles emerging from a neutrino interaction within a given channel, events are often classified according to their \textit{topologies} based on the hadron multiplicity in the final state. We refer to these as \textit{exclusive} topologies, and they are often targeted for cross-section measurements. 
The application of a CC-inclusive-based reconstruction approach introduces a bias in exclusive sub-samples, depending on the difference in proton and pion multiplicities compared to the CC-inclusive average. We thus consider a second scenario where individual response matrices are prepared and applied separately for charged current interactions without charged pions (CC$0\pi$) and those with a single charged pion (CC$1\pi^\pm$), while all other topologies continue to use CC-inclusive matrices. This approach is referred to as \textit{topology-specific calorimetry} in the following, and corresponding lookup tables are shown in Figure~\ref{fig:topolookup_avg_refgen} in the Appendix.

\begin{figure*}[t]
\centering
    \includegraphics[height=0.26\textheight,clip,trim={0 0 4.7cm 0}]{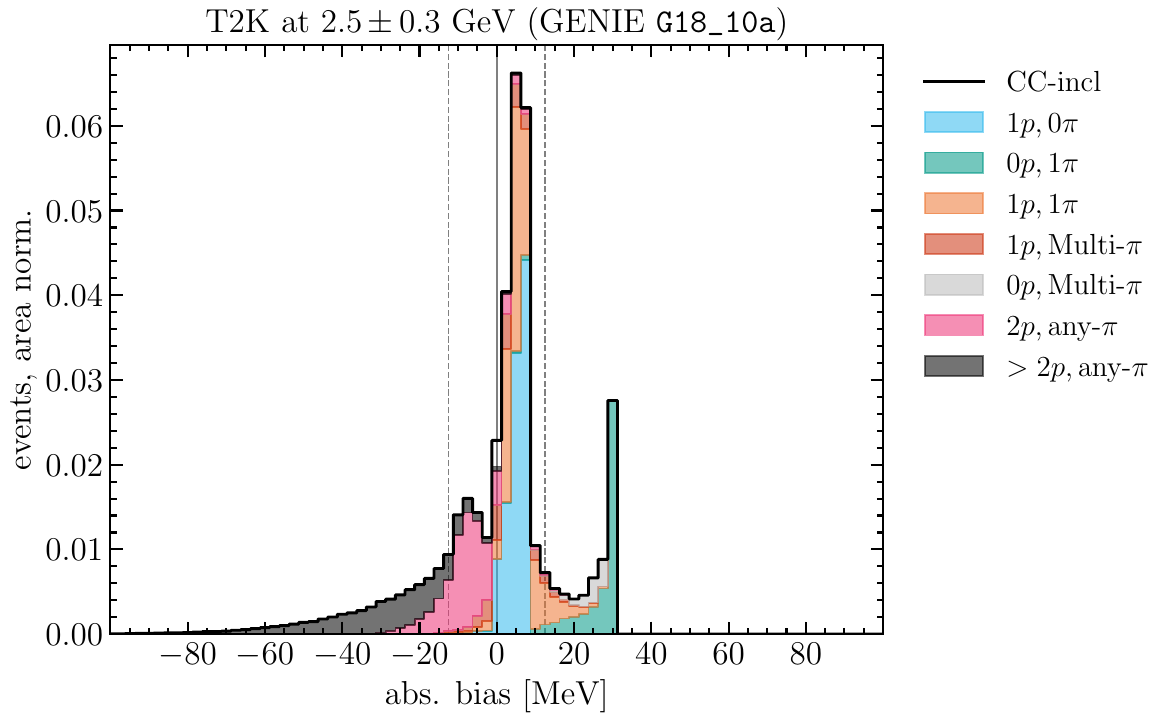}
    \includegraphics[height=0.26\textheight]{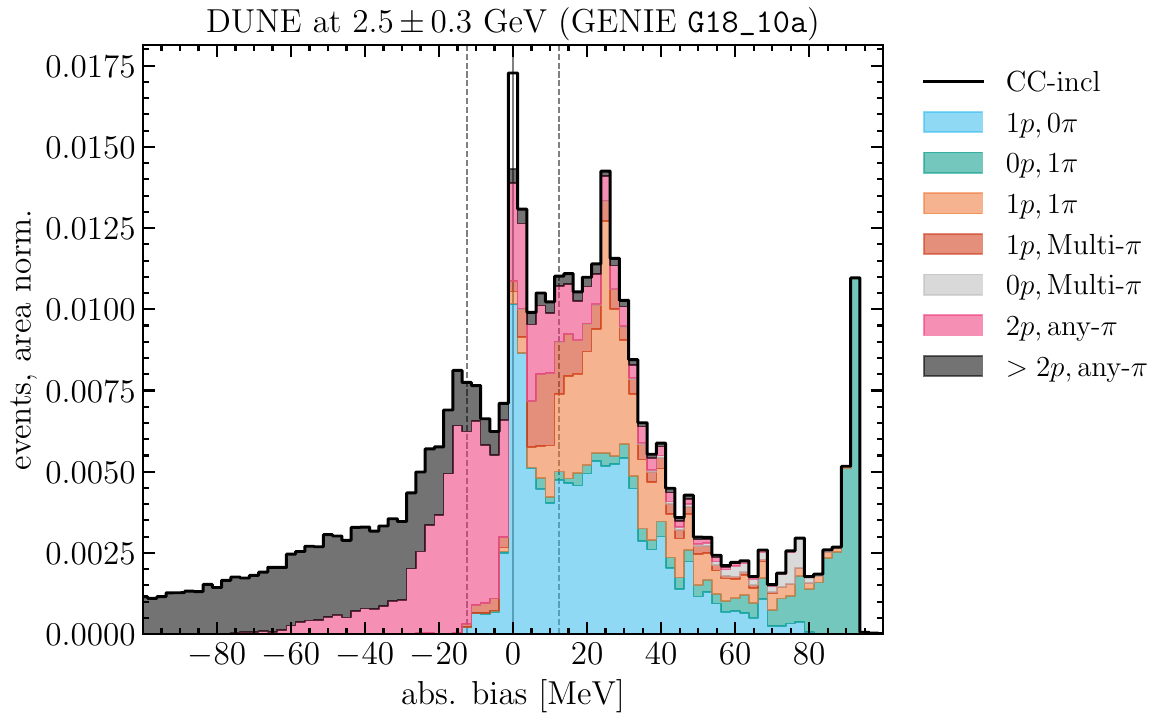}
    \caption{\label{fig:absbias_slices_bynparticle_2500MeV}Bias distributions for the CC-inclusive-based calorimetric approach, showing results for $E_\nu$ slices of~(2.5$\pm$0.3)\,GeV for the T2K ND280 (left) and the DUNE FD (right) for the reference model GENIE~\texttt{G18\_10a}. Colours indicate the composition in terms of hadron multiplicities.}
\vspace{1em}
\centering
    \includegraphics[height=0.27\textheight,clip,trim={0 0 5.5cm 0}]{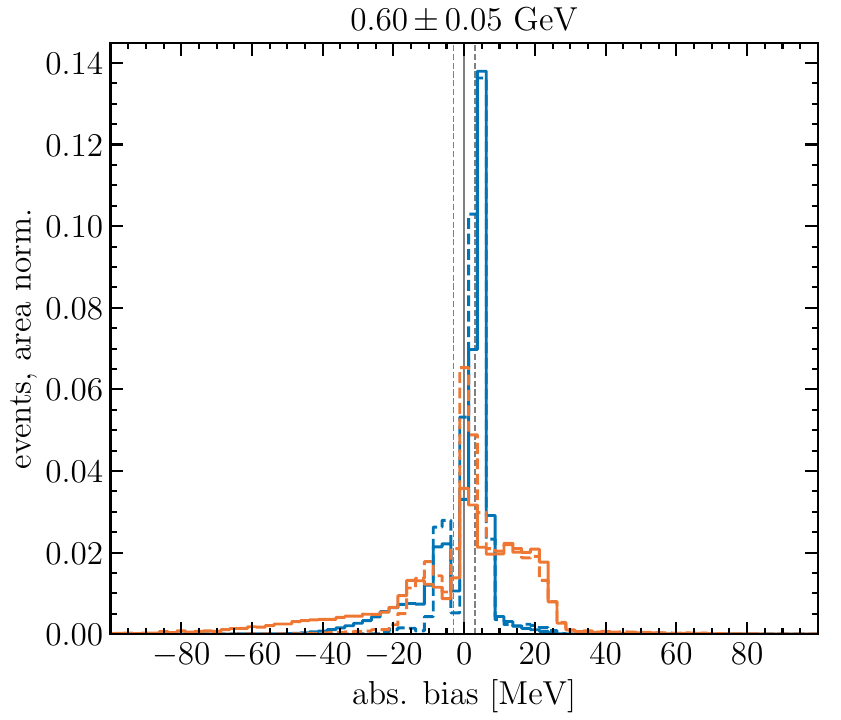}
    \includegraphics[height=0.27\textheight]{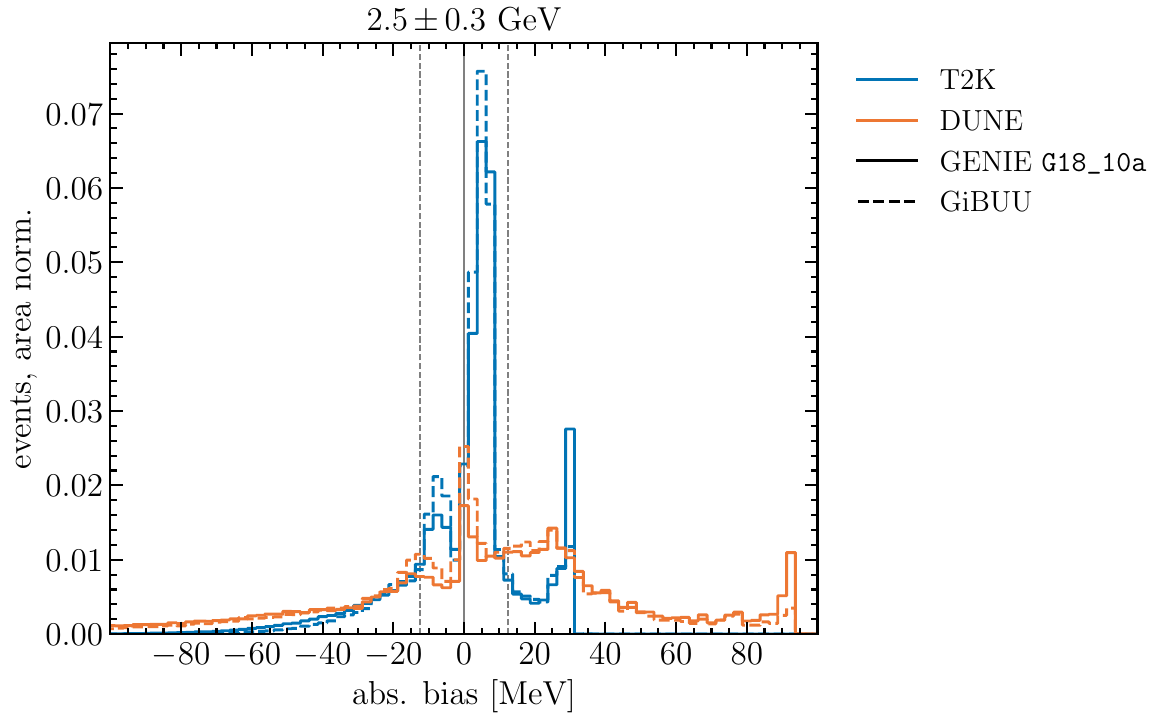}
    \caption{\label{fig:absbias_slices}Bias distributions for the CC-inclusive-based calorimetric approach, showing results for $E_\nu$ slices of~(0.60$\pm$0.05)\,GeV (left) and~(2.50$\pm$0.30)\,GeV (right). Results for the reference model GENIE~\texttt{G18\_10a} (solid lines) are compared to GiBUU (dashed lines) fro the CC-inclusive sample at the T2K ND280 and the DUNE FD.}
\end{figure*}

\subsubsection{Tracking- and hybrid approaches}

Modern detectors such as $\mu$BooNE, DUNE, and the upgraded T2K Near Detector ND280 (in particular its 3D-segmented target, the SuperFGD) provide unprecedented resolution to resolve individual particle tracks.
We consider these tracking capabilities in an alternative reconstruction approach that conservatively assumes all particles above the tracking threshold (Table~\ref{tab:tracking_thresholds}) to be correctly identified, and corresponding energies to be reconstructed without error. This quantity can be used as an estimate for a purely kinematic alternative of $E_\mathrm{had,vis}$, which is missing contributions from vertex activity. The corresponding observable for kinematically reconstructed visible hadronic energy, $E_\mathrm{had,vis}^{reco,\,kin.}$, is defined as the sum of true energies for all \textit{tracked} particles:

\begin{equation}
\label{eq:ehadvis_va_reco_kin}
E_\mathrm{had,vis}^{reco,\,kin.} = \Sigma T^{true}_{p, track} + \Sigma T^{true}_{\pi^\pm, track} + \Sigma E_{\pi^0,\gamma,\mathrm{other}}^{true}.
\end{equation}

The difference between this quantity and $E_\mathrm{had,vis}^{true}$ is only the missing contribution from charged particles in vertex activity.
As vertex activity is still visible as blobs of visible charge around the interaction vertex, it can be added using the calorimetric approach, which defines what is here referred to as \textit{hybrid} reconstruction. For this approach, dedicated response matrices are prepared using only untracked particles, again separately for CC$0\pi$ and CC1$\pi^\pm$ as for the topology-specific calorimetric approach. The calorimetrically reconstructed sum of visible energies in vertex activity is then added to the kinematic result:

\begin{equation}
\label{eq:ehadvis_va_reco_hyb}
E_\mathrm{had,vis}^{reco,\,hyb.} =  E_\mathrm{had,vis}^{reco,\,kin.} + \Sigma T^{reco,topo}_\mathrm{VA} (Q_\mathrm{VA}).
\end{equation}

Regardless of reconstruction approach for $E_\mathrm{had,vis}$, it can be used to obtain a proxy for the neutrino energy by adding the energy of the lepton:

\begin{equation}\label{eq:enu_reco}
    E_\nu^{reco} = E_\mathrm{had,vis}^{reco} + E_\mathrm{lep}.
\end{equation}

As a result, any bias on $E_\mathrm{had,vis}^{reco}$ directly propagates as bias on the reconstructed neutrino energy.

\section{\label{sec:results}Results}

The bias on the visible hadronic energy is defined as: 

\begin{equation}
    \mathrm{abs.\,bias} = E_\mathrm{had,vis}^{reco} - E_\mathrm{had,vis}^{true}.
\end{equation}

This bias quantifies the impact of material effects, while being insensitive to invisible energy due to neutrons or pion masses. Figure~\ref{fig:absbias_slices_bynparticle_2500MeV} shows the bias distribution for the reference model GENIE \texttt{G18\_10a} at T2K and DUNE for a fixed slice of (2.50$\pm$0.30)\,GeV neutrino energy, corresponding to the oscillation maximum at DUNE. The same plots are shown for a second reference value corresponding to the oscillation maximum at T2K/Hyper-K of (0.60$\pm$0.05)\,GeV in
Figure~\ref{fig:absbias_slices_bynparticle_600MeV} in the Appendix. The different colours indicate the contributions of events with distinct hadron multiplicities. The higher $\diff E/\diff x$ of protons compared to pions results in a lower visible charge created at a given particle energy due to the stronger impact of material effects. This results in a strong negative bias for events with multiple protons in the final state, while events with one or less protons tend to be positively biased as the average amount of quenching exceeds that of a single proton, an effect stronger pronounced at DUNE than at T2K. The largest positive bias occurs in topologies without protons and a single charged pion which represents the sample with overall lowest contributing stopping powers, and is thus least affected by material effects.
Low-energetic pions are less affected as the Bragg peak contributes relatively highly to the total energy deposit.
As the pion energy increases, the contribution of $\diff E/\diff x$ at the level of minimal ionisation increases as well, eventually leading to an accumulation of events at a constant bias when the pion energy enters the linear region in the detector response, as can be seen in the horizontal offset between single pion response and the average calorimetric response for the example of T2K shown in Figure~\ref{fig:lookup_single_vs_avg_t2k} in the Appendix.

Figure~\ref{fig:absbias_slices} shows bias distributions for the CC-inclusive sample at T2K and DUNE for both reference energies of (0.60$\pm$0.05)\,GeV and (2.50$\pm$0.30)\,GeV for the reference model as well as GiBUU as an alternative model. The relative abundance of different topologies and corresponding particle multiplicities changes with the neutrino energy, resulting in the difference in shape between the two plots. The comparison to GiBUU highlights a strong model dependence in particular in the peaks dominated by $1p$ and $2p$ final states.

To visualise the bias evolution with the neutrino energy, we evaluate the mean bias across all CC-inclusive events:

\begin{equation}
    \langle  \mathrm{abs.\,bias} \rangle = \frac{1}{N_\mathrm{evts}}\sum_{\mathrm{evts}} \left( E_\mathrm{had,vis}^{reco} - E_\mathrm{had,vis}^{true}\right).
\end{equation}

\begin{figure*}[t]
  \includegraphics[width=\textwidth]{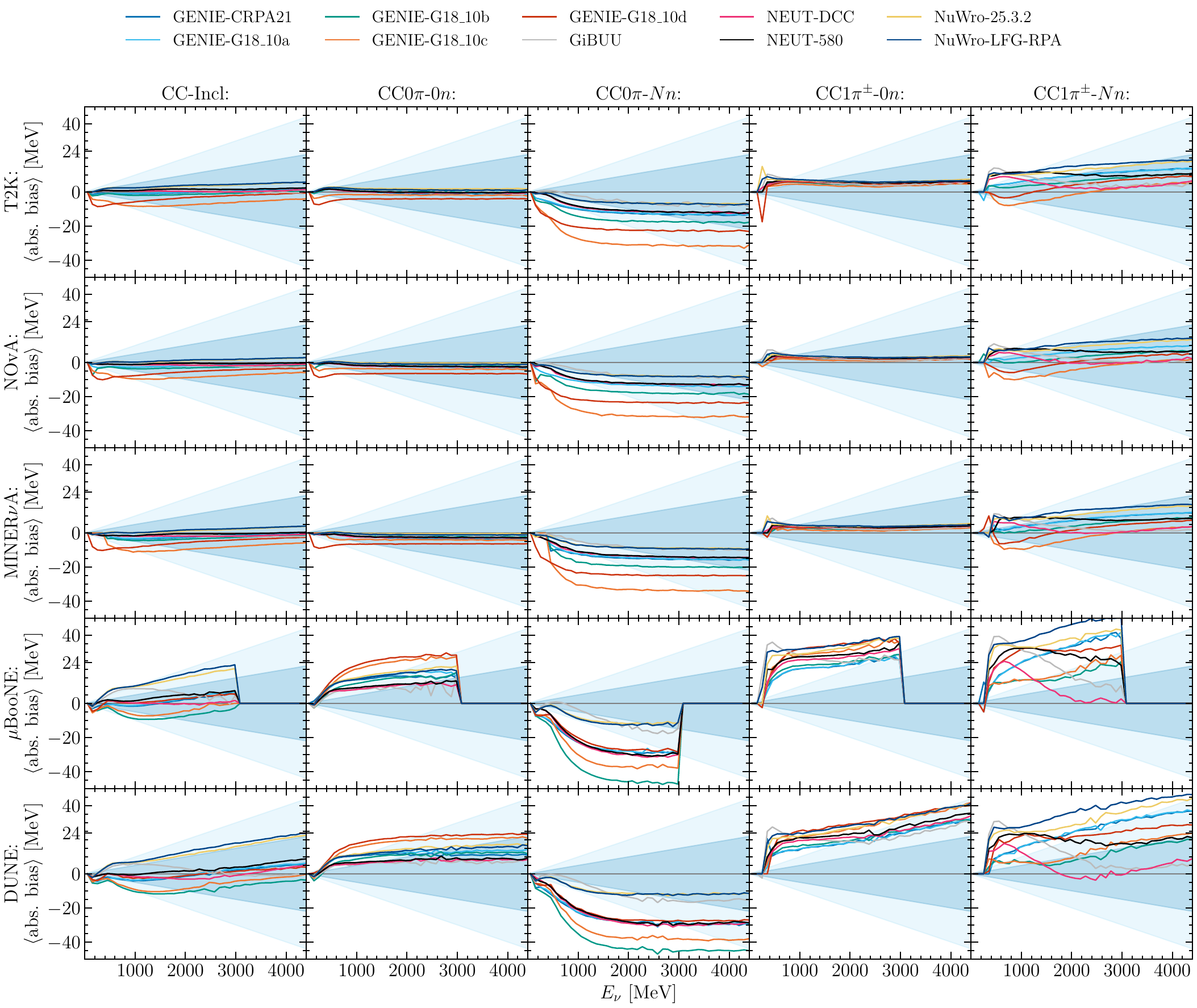}
  \caption{\label{fig:absbias_vs_enu} 
    Mean bias caused by material effects as a function of the true neutrino energy $E_\nu$ for the CC-incl.-based calorimetric reconstruction approach. Results are shown for T2K, NO$\nu$A, MINER$\nu$A, $\mu$BooNE (simulated up to 3\,GeV), and
    DUNE (from top to bottom), for different interaction
    topologies (columns). Dark (light) shaded regions indicate $\pm$0.5\% ($\pm$1\%) bias on $E_\nu$. 
    }
\end{figure*}

\begin{figure*}[t]
    \begin{center}
    \includegraphics[height=0.45\textheight,trim={0 0 0 5.5mm},clip]{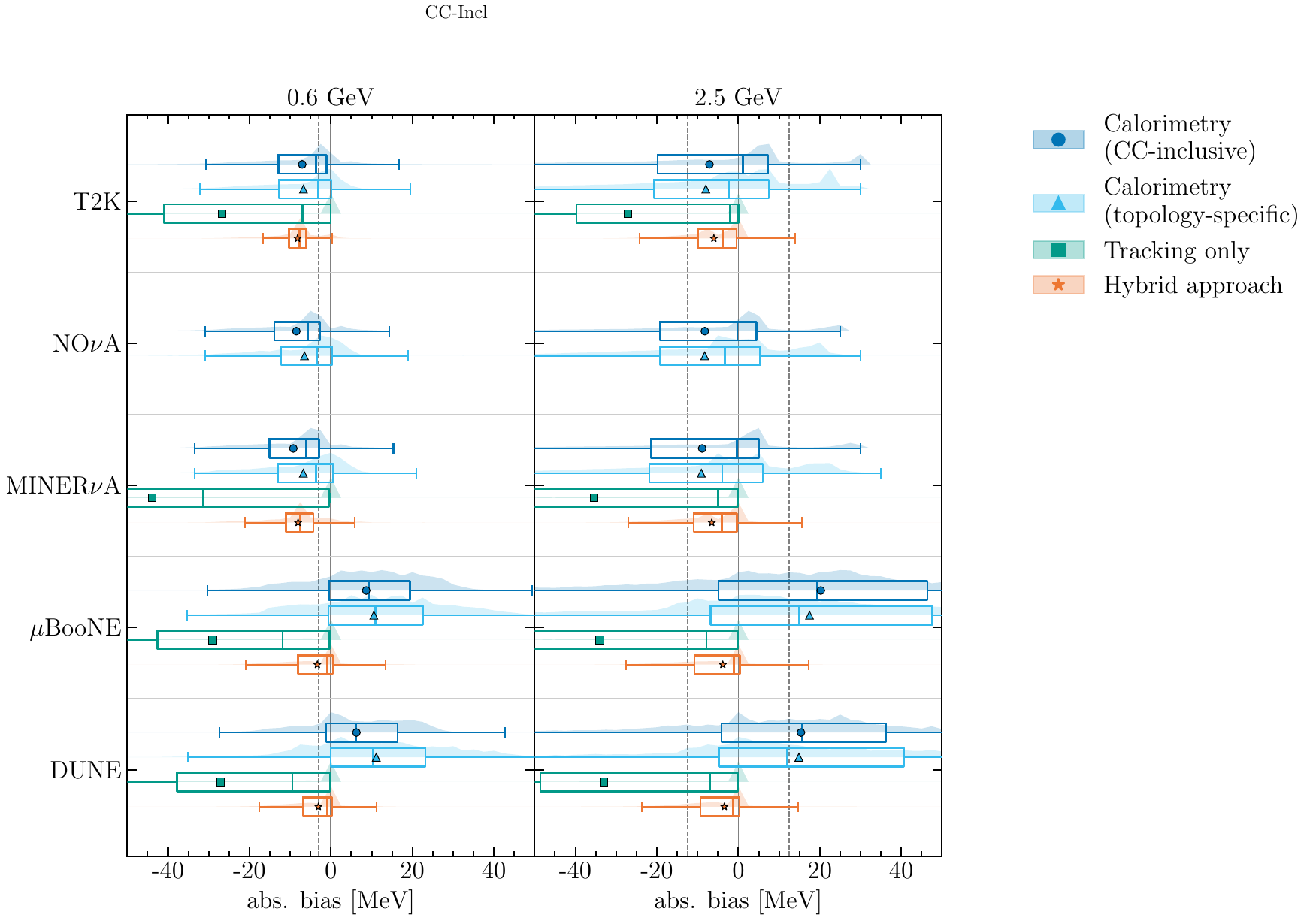}
    \end{center}
    \caption{Bias distributions for slices of $E_\nu^{true}$ of (0.60$\pm$0.05)\,GeV and (2.50$\pm$0.05)\,GeV in the CC-inclusive sample, showing mean bias (marker), box plots indicating the central 50\% of events (interquartile range, IQR) with the median as vertical line and whiskers to $1.5\times$IQR. Shaded histograms show the full bias distribution. Vertical dashed lines indicate $\pm$0.5\% of the given value of $E_\nu$.}
    \label{fig:boxplots_absbias}
\end{figure*}

Results are shown as a function of the true neutrino energy in Figure~\ref{fig:absbias_vs_enu} for different interaction models when applying the CC-inclusive-based calorimetric method. Results are shown for the CC-inclusive as well as exclusive samples of CC$0\pi$ and CC$1\pi$ topologies, which are further split into subsamples depending on the presence of at least one neutron ($0n$ and $Nn$). Shaded regions indicate a $\pm$0.5\% (dark blue) and $\pm$1\% (light blue) deviation from the true value of $E_\nu$. In general, largest mean biases are observed in exclusive topologies with neutrons in the final state, as the average energy available to charged hadrons is lower. Topologies with no charged pion and neutrons in the final state generally exhibit negative biases at values of around -30\,MeV for experiments using scintillators, and -45 to -50\,MeV for LAr-TPCs. The mean biase remains constant at these values from 1\,GeV up to higher neutrino energies. CC$1\pi^\pm$ events show a positive bias, with some exceptions in the CC$1\pi^\pm Nn$ sample, which overall shows the largest variation between models. Here, mean biases of around 15-20\,MeV for scintillator-based experiments and around 50\,MeV for the LAr-TPCs are observed. It should be noted that $\mu$BooNE events are only simulated up to $E_\nu$ of 3\,GeV. Analogous results for the hybrid reconstruction approach can be found in the Appendix (Figure~\ref{fig:absbias_hybrid_vs_enu}), and are overall significantly lower. Interaction models with INC-based FSI (GENIE \texttt{G18\_10c} and \texttt{G18\_10d}) exhibit overall significantly stronger biases in the hybrid approach than models without, although not exceeding -15\,MeV. Results for the other methods, as well as versions showing the mean bias in dependence on the visible hadronic energy, are provided in the supplementary material.

Next, bias distributions are contrasted across the four reconstruction methods described above, namely the pure calorimetric approach with a CC-inclusive-based smearing matrix (\textit{CC-incl.-based calorimetry)}, the version with dedicated response matrices for CC$0\pi$ and CC$1\pi$ events \textit{topology-specific calorimetry)}, the purely \textit{tracking-based approach} that neglects contributions from vertex activity ($E_\mathrm{had,vis}^{reco,\,kin.}$), and finally the \textit{hybrid approach} ($E_\mathrm{had,vis}^{reco,\,hyb.}$). 
Figure~\ref{fig:boxplots_absbias} shows bias distributions for the interaction model that exhibited the highest mean bias at these reference values, comparing all four reconstruction approaches. Box plots indicate median and the central 50\% of events (interquartile range, IQR), and the whiskers extend to $1.5\times$IQR corresponding to the range of 75\% of events. Markers indicate the mean bias, and Table~\ref{tab:results_absbias} in the Appendix provides the corresponding values of mean biases. The full bias distribution for the given slice of neutrino energy is shown as a shaded histogram in the background. While mean biases obtained by the hybrid approach remain within 0.5\% at 2.5\,GeV neutrino energy for all experiments, no experiment meets the target at 0.6\,GeV.

For the hybrid approach at DUNE, 28.1\% of all CC-inclusive events in the 2.5\,GeV still show a bias exceeding 12.5\,MeV (0.5\% of $E_\nu$), and corresponding results of the fraction of events exceeding the 0.5\% limit are summarised for all experiments in Table~\ref{tab:frac_evts_above_thresh} in the Appendix. In exclusive samples without neutrons in the final state, this number is reduced to around 10\% of all events at 2.5\,GeV due to the overall higher energies transferred to charged hadrons, which result in a lower relative contribution of quenching. In topologies with neutrons on the other hand, 20-30\% of CC$1\pi^\pm Nn$ and 35-46\% of CC$0\pi^\pm Nn$ events exceed  0.5\% bias at this energy. This increase is caused by the overall lower energy of hadrons when a fraction of the energy is carried away by neutrons, leading to a comparatively higher impact of material effects. 

\section{\label{sec:discussion}Discussion}

Material effects such as Birks quenching for scintillators and recombination effects in LAr-TPCs have a non-negligible impact on calorimetrically reconstructed visible hadronic energies, and consequently, neutrino energies. Smearing matrices for calorimetric $E_\mathrm{had,vis}$ as shown in Figure~\ref{fig:ehadvis_smear} indicate that recombination effects in LAr-TPCs, $\mu$BooNE and DUNE-FD, generally have a stronger impact than than quenching in scintillator-based detectors, T2K's ND280, MINER$\nu$A, and NO$\nu$A, which was confirmed by resulting biases. 

Figure~\ref{fig:absbias_vs_enu} indicates that  for the CC-inclusive-based calorimetric approach material effects contribute a mean bias exceeding 0.5\% of $E_\nu$ below 2\,GeV for all examined experiments, which for DUNE remains true over entire evaluated range of up to 4\,GeV. Topology-specific tuning of the reconstruction can even increase the bias due to stronger variation in model predictions for certain topologies (see for example CC$0\pi$-$0n$ and $Nn$ at DUNE), as Figure~\ref{fig:boxplots_absbias_topo} shows. Exploiting the tracking capabilities provided by modern neutrino detectors is thus desirable.

However, results for the CC-inclusive sample shown in Figure~\ref{fig:boxplots_absbias} further demonstrate that even perfect tracking without error above the published thresholds leads to significant negative biases due to the lack of inclusion of charged particles in vertex activity. The hybrid approach mitigates this by adding a calorimetric sum of untracked particle energies, and was found to be the only method for LAr-TPCs ($\mu$BooNE and DUNE) that yields mean biases below 0.5\% of $E_\nu$ at 2.5\,GeV. As can be inferred fromTable~\ref{tab:frac_evts_above_thresh}, 28.1\% of events for DUNE were still found to contribute a bias greater than 0.5\% at this reference energy value, with even greater percentages at 0.6\,GeV neutrino energy, the latter potentially strongly impacting measurements of the second oscillation maximum around 1\,GeV. The SBND~\cite{SBND:2025lha} experiment is placed to detect the same beam composition as $\mu$BooNE, however uses a higher electric field at the same value as DUNE, which reduces recombination effects. Biases can therefore be expected to smaller than here evaluated for $\mu$BooNE.

These findings highlight the importance of dedicated cross-section measurements at detectors capable of resolve particle tracks in what detectors discussed here only detect as vertex activity. Examples of detector technologies capable of providing such measurements include trackers made of scintillating fibres~\cite{Hyper-Kamiokande:2025asb}, high-pressure TPCs~\cite{DUNE:2021tad}, or emulsion-based designs~\cite{blau1937disintegration, Fukuda:2017clt}. 

It has to be noted that all the quantitative results of this study are highly model dependent. The bias evaluation relies on the assumption of GENIE \texttt{G18\_10a} as the reference model - a different choice of reference model likely yields different results. Results for GENIE \texttt{G18\_10c} and  \texttt{G18\_10d} that predict the liberation of nuclear clusters are underestimated due to the treatment of all heavy nuclear clusters as alpha particles for simplicity (see Section~\ref{sec:sim}), and the impact of material effects is in reality stronger for heavier particles. Moreover, while being a conservative assumption, the hybrid method is based on error-free particle identification and energy reconstruction above the tracking threshold, which is unrealistic in an actual experiment, and mis-reconstruction of particle tracks introduces additional bias. Electromagnetic showers as for example caused by neutral pions are not subjected to material effects in this study either. The biases presented here likewise exclude those arising from corrections for pion masses and the missing energy of neutral particles, which are more commonly considered in literature.
Results should thus be interpreted as a conservative minimum, quantifying solely the impact of aforementioned material effects, rather than a complete bias evaluation for calorimetric $E_\nu$.

Overall, it can be concluded that interaction-model-dependent biases introduced by non-linearities in the detector response introduce significant systematic uncertainties inherent to calorimetric reconstruction approaches. While detectors discussed here represent state-of-the-art trackers, energy deposits from untracked particles in vertex activity remain a challenge to incorporate in neutrino energy reconstruction. This is compounded by the fact that modelling uncertainties are particularly pronounced for interactions with low energy transfer, which is especially prone to producing vertex activity. Oscillation analyses as well as cross-section measurements thus have to attribute sufficient systematic uncertainties to cover this effect.

As long as no improved external inputs exist to provide stronger model constraints, only a change in the analysis strategy could mitigate these inherent biases. A suggestion for such an alternative strategy is described in the next section.

\section{\label{sec:alt_appr}Alternative analysis strategies}

Data analysis approaches in neutrino cross-section measurements are typically based on a deconvolution of detector effects and inefficiencies from experimental data, with the goal to extract information from unfolded ``truth-level'' quantities. 
The fundamental limitation of such an approach, as demonstrated in this paper, is that multiple true states can produce the same observed state, and the mapping between the two is highly model dependent. Not only the material effects discussed above, but also other detector effects for example in electronics, as well as reconstruction and selection inefficiencies, can cause similar biases. 

An alternative strategy is to change the direction in the flow of information by applying such corrections via response matrices on simulated models rather than the measured data. This method, referred to as forward folding, has been suggested to reduce bias in interaction cross-section measurements~\cite{Avanzini:2021qlx, dolan2026cpviolation,koch2019response, gardiner2024mathematical}. 

The MiniBooNE and $\mu$BooNE collaboration have explored this technique~\cite{MiniBooNE:2010xqw, MicroBooNE:2019nio, MicroBooNE:2020akw}, although similar biases in the detector response matrices applied in these cross-section measurements were reported~\cite{Avanzini:2021qlx}. Insufficient dimensionality in response matrices for forward folding causes the same problem as an unfolding-based analysis approach, as can also be seen in this study: Response matrices for $E_\mathrm{had,vis}$ are inherently model-dependent, and therefore introduce a bias irrespective of the direction in which they are applied. This also affects fits of oscillation parameters when using response matrices based on variables that include calorimetric sums of hadron energies, leading to the aforementioned biases.

This problem can be resolved by increased dimensionality in the response matrix. In particular, instead of applying response matrices on a per-event basis, responses can be applied on each simulated particle individually. Detectors with tracking capabilities typically use control samples to tune their simulations of individual particles, which would even allow data-driven inputs to per-particle response matrices. This information could be published alongside experimental data, which could then be released even closer to the detector level rather than in reconstructed observables, that is, in distributions of $\Sigma Q_\mathrm{had,vis}$ instead of $\Sigma E_\mathrm{had,vis}^{reco}$ for a differential cross-section measurement in visible hadronic charge rather than energy. Individual particle kinematics are always known at truth, which allows to work around the problematic ambiguities that arise otherwise. 

\begin{figure}[t]
    \includegraphics[width=.9\linewidth]{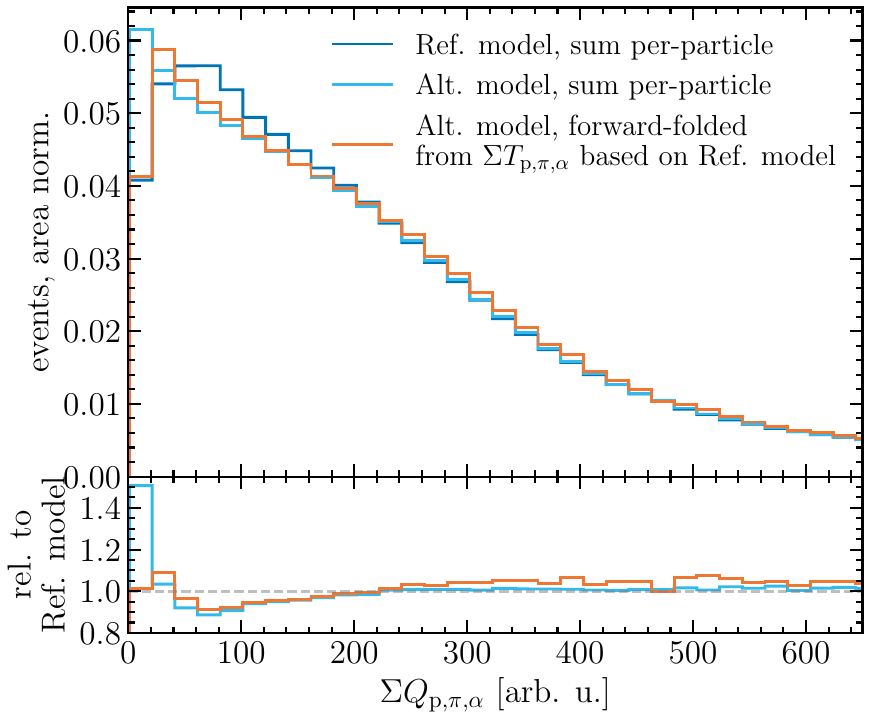}
    \caption{Comparison of the sum of visible charges from protons, charged pions, and nuclear clusters ($\Sigma Q_{p,\pi,\alpha}$ in the reference model GENIE~\texttt{G18\_10a} (dark blue) for the T2K ND280, from which a one-dimensional $\Sigma T_{p,\pi,\alpha}\leftrightarrow\Sigma Q_{p,\pi,\alpha}$ response matrix is built, the forward folded spectrum of GENIE~\texttt{G18\_10c} using this matrix (orange), as well as using a per-particle forward folding approach (light blue). Bottom sub-plot shows the ratio to the reference model.
    \label{fig:forward_fold}
    }
\end{figure}

Figure~\ref{fig:forward_fold} illustrates this idea: the reconstruction based on the reference model GENIE \texttt{G18\_10a} for T2K is applied in reverse direction to obtain a one-dimensionally forward-folded result of $\Sigma Q_{p,\pi,\alpha}$ from the true value of $\Sigma T_{p,\pi,\alpha}$ in an alternative model. The resulting distribution of visible charges is compared to that of the reference model itself, as well as the distribution obtained when calculating the sum of visible charges for each individual particle, that is, applying the per-particle response approach. A clear difference in the forward folded spectrum can be observed, as highlighted by the ratio of events relative to the reference model, and at the low end of the spectrum the one-dimensionally forward folded spectrum replicates the reference model more than its true shape.

However, new challenges come with such a per-particle-response approach: backgrounds have to be quantified and added in a compatible and model-independent way, as do overall efficiency corrections. Other corrections such as particle tracking thresholds are more complex than simple step functions. The publication of robust response matrices thus requires great care to ensure biases are not introduced at a different stage.

A further challenge of such a per-particle-response-based forward-folding approach and the presentation of results at the detector level lies in the communication to model builders and theorists. Comparisons between results from different experiments are not straightforward, as not a single number such as a total cross-section can be quoted. A solution could be an interactive cross-experiment framework that allows to access underlying truth information for a given value of a detector-level observable, and to compare to data published by multiple experiments in parallel. This however is only applicable to interaction models that provide modelling of hadronisation and FSI to predict individual particle kinematics.

While such strategies are mostly discussed in the context of cross-section measurements, similar methods can be applied in oscillation analyses as well. The decisive advantage of a forward-folding approach based on per-particle response matrices is that it follows the direction in which information flows unambiguously: from truth to observation. No inversion is attempted, and hence no model-dependent assumptions enter the construction of any quantity involved. 

\section{Conclusion} 

Next generation neutrino oscillation experiments Hyper-K and DUNE heavily rely on improvements in neutrino interaction modelling. Theoretical predictions differ particularly strongly for interactions at low energy transfer to the nucleus. This typically results in the liberation of particles below the tracking thresholds of modern neutrino detectors, making benchmarking of such models a challenge. Calorimetrically reconstructed observables provide a handle on energy deposits in untracked particles, but the impact of non-linear effects in the detector material, in particular Birks quenching and ionisation charge recombination, was found to introduce significant interaction-model-dependent biases on the reconstructed visible hadronic energy, and through this quantity, the neutrino energy. Corresponding systematic uncertainties have to be assigned to such variables to cover these effects. Detectors of improved tracking resolution compared to current state-of-the-art detectors such as SuperFGD of T2K's ND280, the DUNE Far Detector, or $\mu$BooNE, are required to resolve ambiguities in energy deposits from untracked particles. Forward folding does not necessarily present a viable alternative when response matrices cover insufficiently many dimensions. A forward-folding-based analysis approach built on per-particle response matrices, however, could circumvent the problem of insufficient dimensionality in the response.

\begin{acknowledgments}

The authors would like to thank Callum Wilkinson for sharing NUISANCE flat trees for interaction models in ~\cite{dolan2026cpviolation}. In addition, K. L. would like to thank Tomislav Vladisavljevic for first mentioning Birks quenching in the context of a T2K cross-section measurement at ND280.

\end{acknowledgments}

\section*{Data availability}
All files were generated with the open-source code framework NUISANCE, using official flux files provided by each of the experiments, and can be found at \href{https://portal.nersc.gov/project/nuisance/ MC_IOP_review/}{\texttt{https://portal.nersc.gov/project/nuisance/ MC\_IOP\_review/}} and for GENIE \texttt{G18\_10d} at \href{https://cernbox.cern.ch/s/fpfJPOAPaSdd7h5}{\texttt{https://cernbox.cern.ch/s/fpfJPOAPaSdd7h5}}. The analysis framework developed for this study is available at \href{https://gitlab.com/klachner/birks-quenching-studies/-/tree/main?ref_type=heads}{\texttt{https://gitlab.com/klachner/birks-quenching-studies}}.

\bibliographystyle{apsrev4-2}
\bibliography{references} 

\section*{Appendix}

\appendix

A summary of all interaction models considered in this paper can be found in Tables~\ref{tab:fsi} and \ref{tab:models}, with more details provided in Ref.~\cite{dolan2026cpviolation}.

\begin{figure*}[t]
    \centering
    \includegraphics[width=0.7644\textwidth, clip, trim={0 1.2cm 0 1.2cm}]{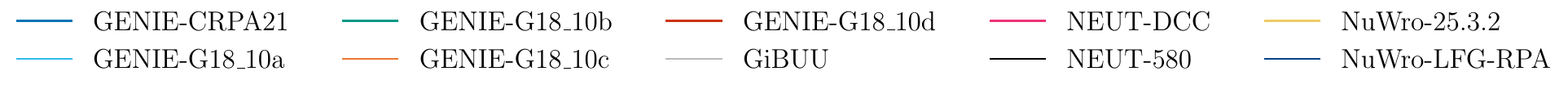}  
    \begin{subfigure}{0.49\textwidth}
    \includegraphics[width=.78\textwidth]{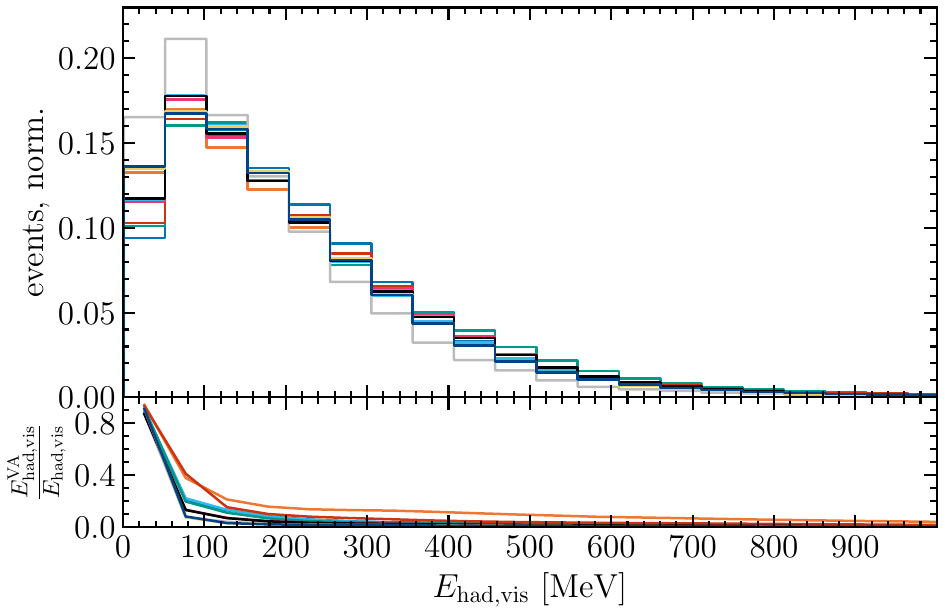}    
    \caption{T2K-ND280}
    \end{subfigure}\hspace{-1.5cm}%
    \begin{subfigure}{0.49\textwidth}
    \includegraphics[width=.78\textwidth]{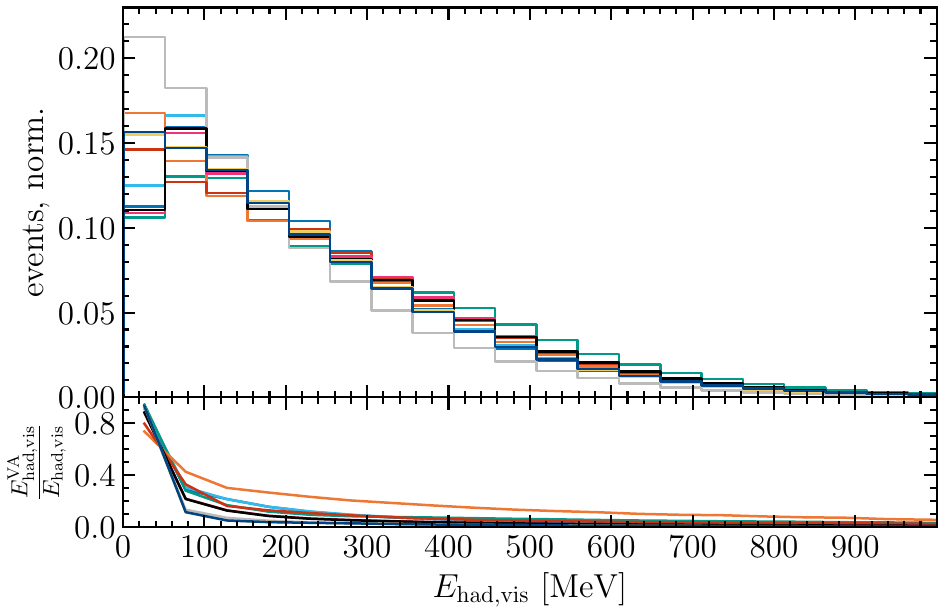}  
    \caption{$\mu$BooNE}
    \end{subfigure}
    \begin{subfigure}{0.49\textwidth}
    \includegraphics[width=.78\textwidth]{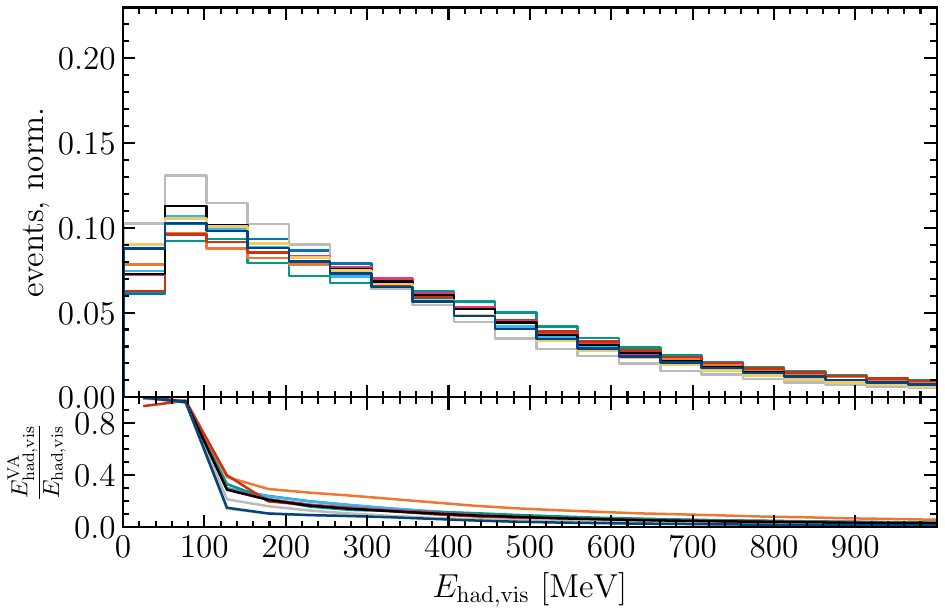}%
    \caption{MINER$\nu$A}
     \end{subfigure}\hspace{-1.5cm}%
    \begin{subfigure}{0.49\textwidth}
    \includegraphics[width=.78\textwidth]{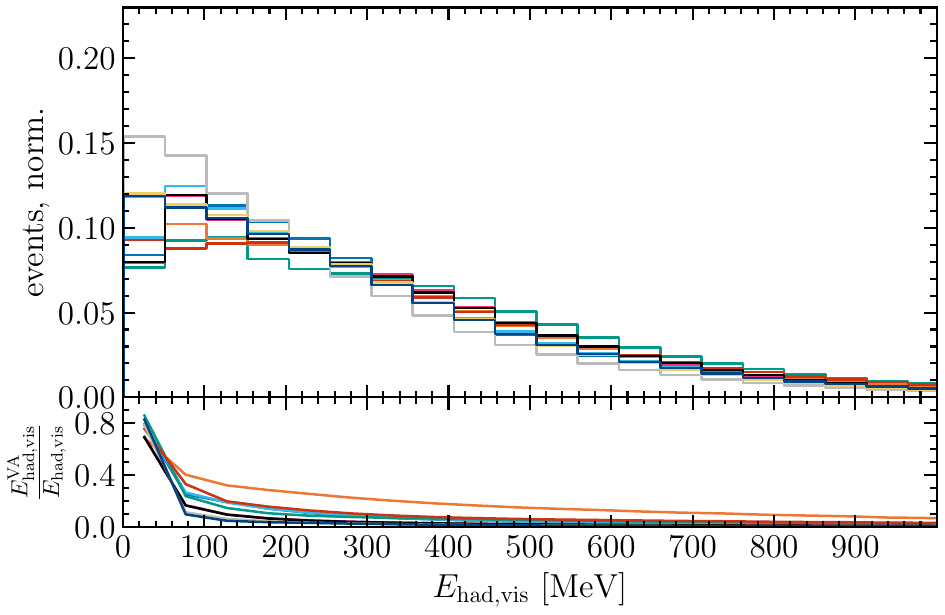} 
    \caption{DUNE}
    \end{subfigure}%
    \caption{\label{fig:ehadvis_cc0pi}Distribution of true $E_\mathrm{had,vis}$ in
    CC$0\pi$ events at T2K (a), $\mu$BooNE
(b), MINER$\nu$A (c), and DUNE (d) for different generators, and fractional contribution of untracked particles (bottom sub-plots).}

\vspace{0.75em}

    \includegraphics[width=0.7644\textwidth, clip, trim={0 1.2cm 0 1.2cm}]{figures/ehadvis_allgen/legend_ehadvis_allgen.pdf}  
    \begin{subfigure}{0.49\textwidth}
    \includegraphics[width=.78\textwidth]{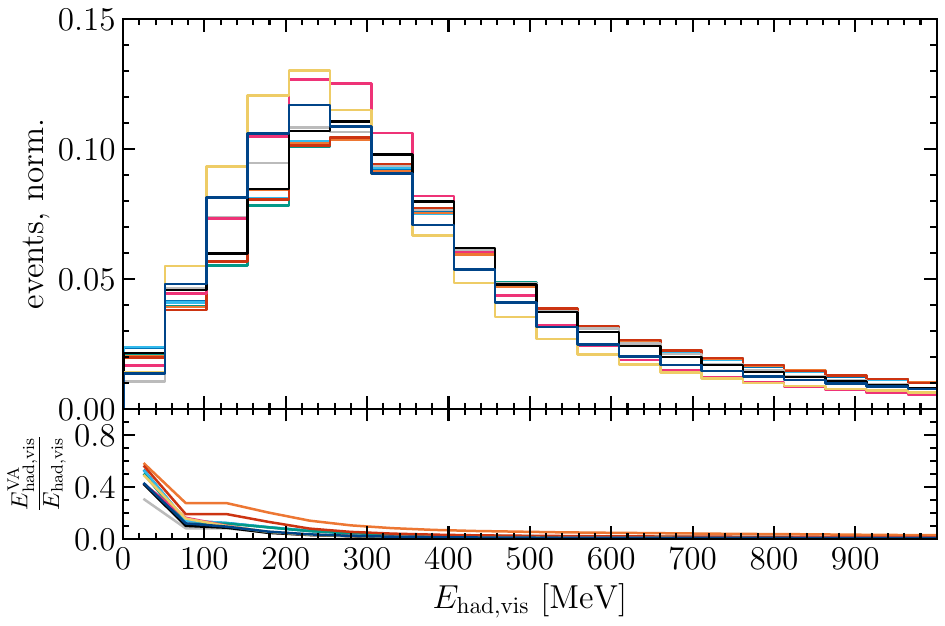}%
    \caption{T2K-ND280}
    \end{subfigure}\hspace{-1.5cm}%
    \begin{subfigure}{0.49\textwidth}
    \includegraphics[width=.78\textwidth]{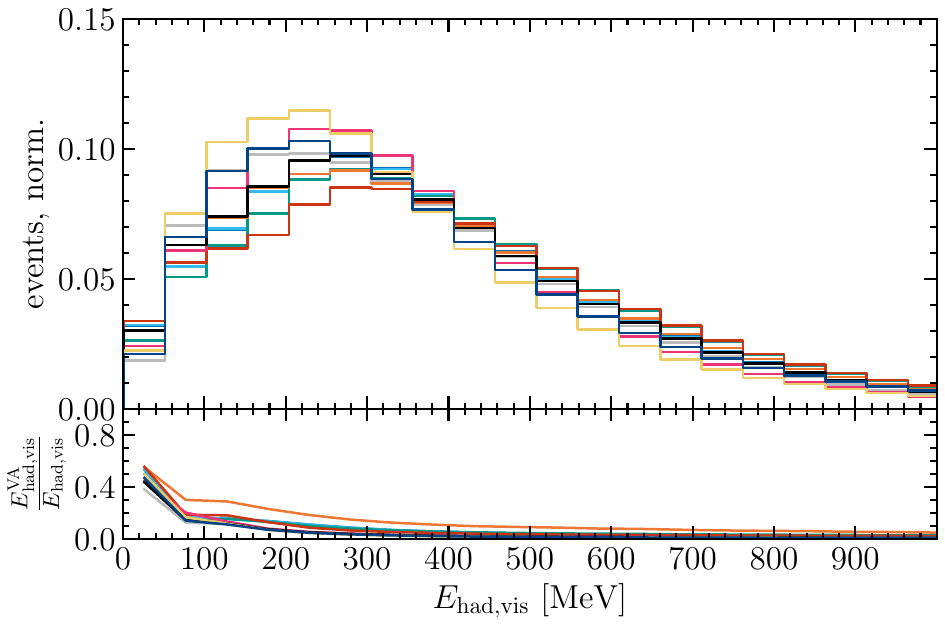} 
    \caption{$\mu$BooNE}
    \end{subfigure}
    \begin{subfigure}{0.49\textwidth}
    \includegraphics[width=.78\textwidth]{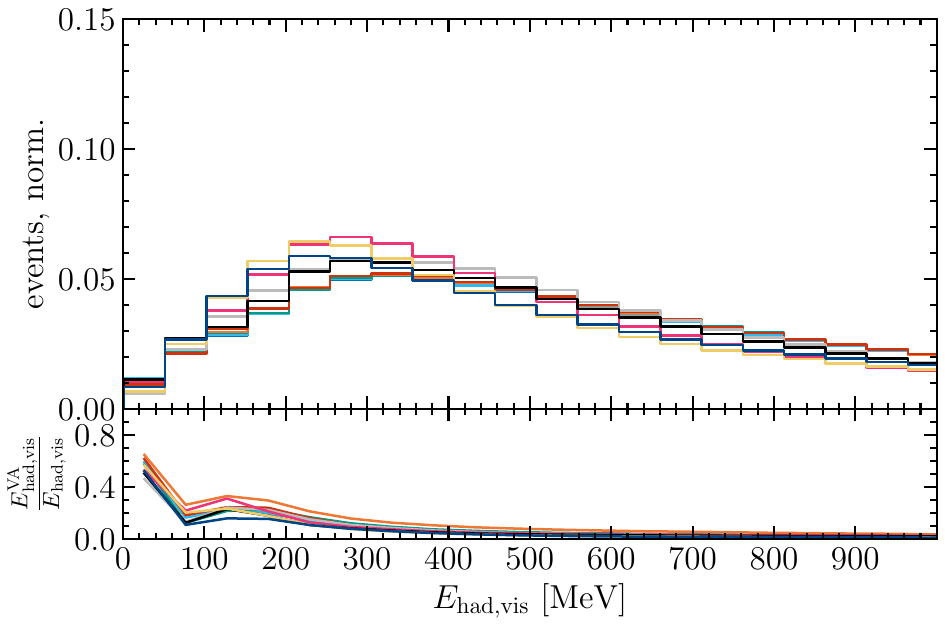}%
    \caption{MINER$\nu$A}
    \end{subfigure}\hspace{-1.5cm}%
    \begin{subfigure}{0.49\textwidth}
    \includegraphics[width=.78\textwidth]{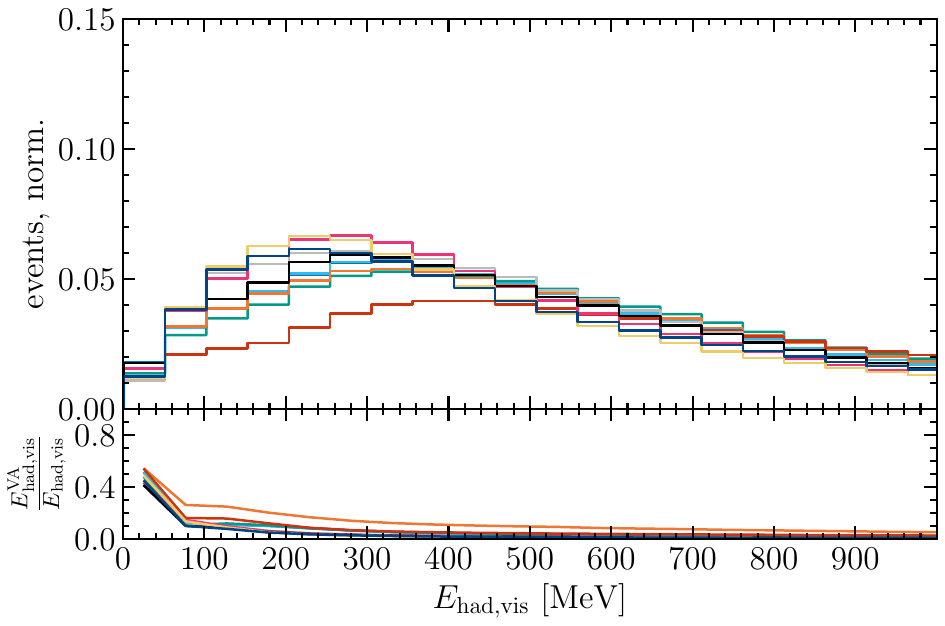}  
    \caption{DUNE}
    \end{subfigure}%
    \caption{\label{fig:ehadvis_cc1pi}Distribution of true $E_\mathrm{had,vis}$ in
    CC$1\pi$ events at T2K (a), $\mu$BooNE
(b), MINER$\nu$A (c), and DUNE (d) for different generators, and fractional contribution of untracked particles (bottom sub-plots).}
\end{figure*}

Figures~\ref{fig:ehadvis_cc0pi} and \ref{fig:ehadvis_cc1pi} show distributions of true $E_\mathrm{had,vis}$ in CC$0\pi$ and CC$1\pi^\pm$ events, respectively. Sub-plots indicate the fraction of energy carried away by untracked particles at the given experiment.

\begin{figure*}
    \centering
    \includegraphics[width=0.6\linewidth]{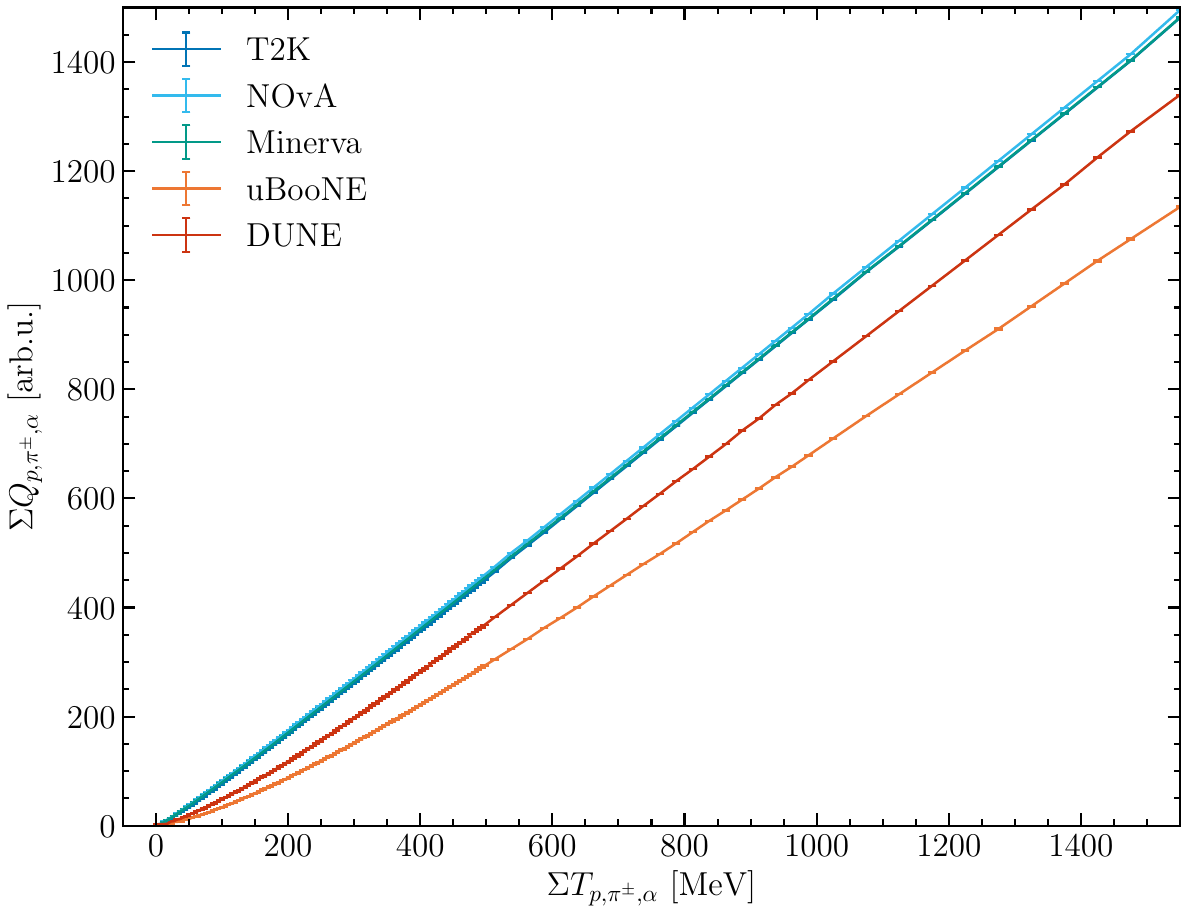}
    \caption{Correspondence between the average sum of kinetic energies of protons, pions, and nuclear clusters ($\Sigma T_{p,\pi^\pm,\alpha}$) and the sum of corresponding quenched light yields ($\Sigma Q_{p,\pi^\pm,\alpha}$) based on the reference model GENIE \texttt{G18\_10a} at all five experiments.}
    \label{fig:lookup_avg_refgen}
\end{figure*}

\begin{figure*}
    \centering
    \includegraphics[width=0.6\linewidth, clip, trim={0 0 0 0.7cm}]{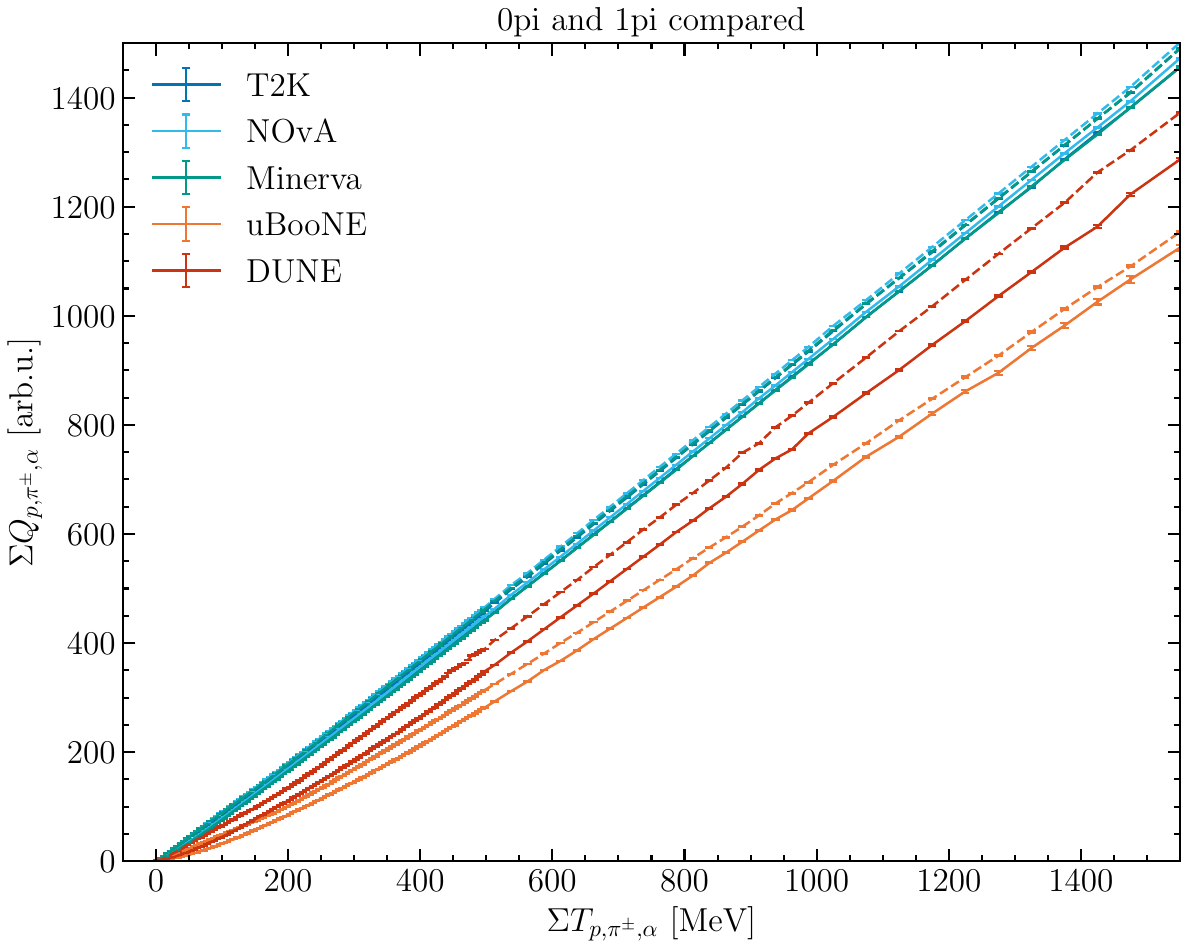}
    \caption{Correspondence between the average sum of kinetic energies of protons, pions, and nuclear clusters ($\Sigma T_{p,\pi^\pm,\alpha}$) and the sum of corresponding quenched light yields ($\Sigma Q_{p,\pi^\pm,\alpha}$) in CC$0\pi$ (solid) and CC$1\pi$ topologies (dashed), based on the reference model GENIE \texttt{G18\_10a} at all five experiments.}
    \label{fig:topolookup_avg_refgen}
\end{figure*}

The response functions describing the dependency of the total sum of visible hadronic charges from protons, charged pions, and nuclear clusters ($\Sigma Q_{p,\pi,\alpha}$), on the average sum of corresponding kinetic energies ($\Sigma T_{p,\pi,\alpha}$) are shown in Figure~\ref{fig:lookup_avg_refgen} for the CC-inclusive-based approach at all experiments, and in Figure~\ref{fig:topolookup_avg_refgen} for the topology-specific approach, with solid lines showing those for CC$0\pi$, and dashed lines those for CC$1\pi$ topologies.

\begin{figure*}
    \centering
    \includegraphics[height=0.33\textwidth]{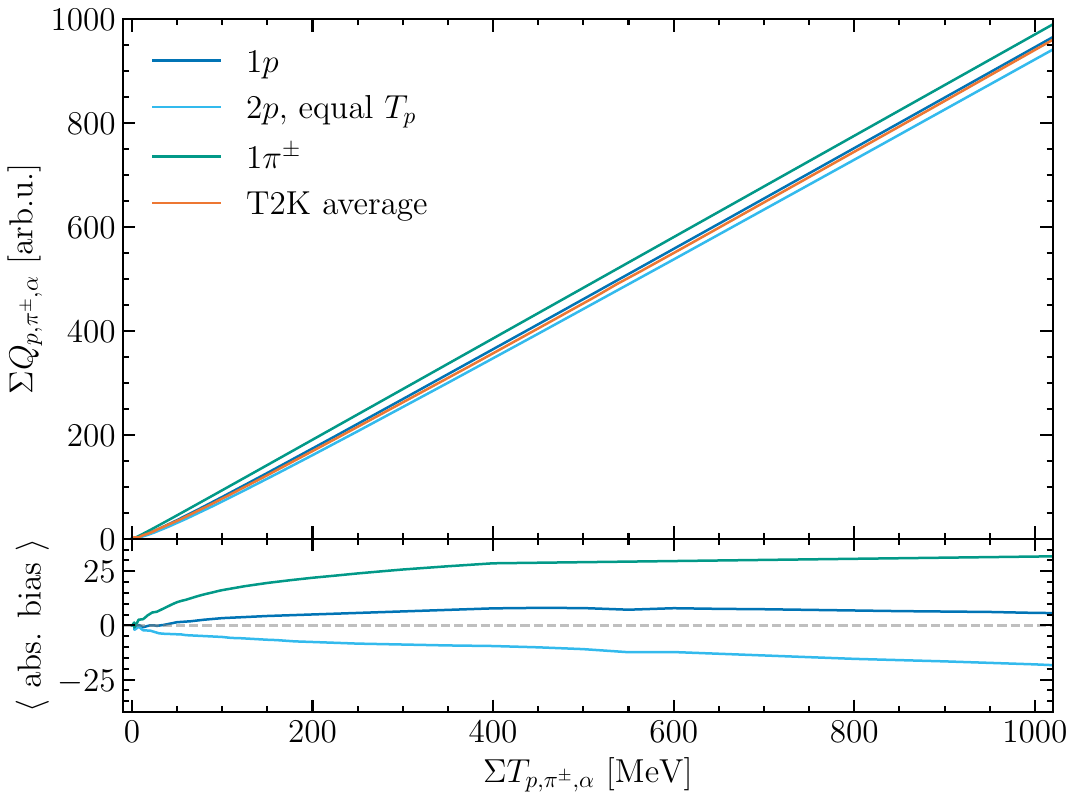}
    \hspace{1cm}
    \includegraphics[height=0.33\textwidth]{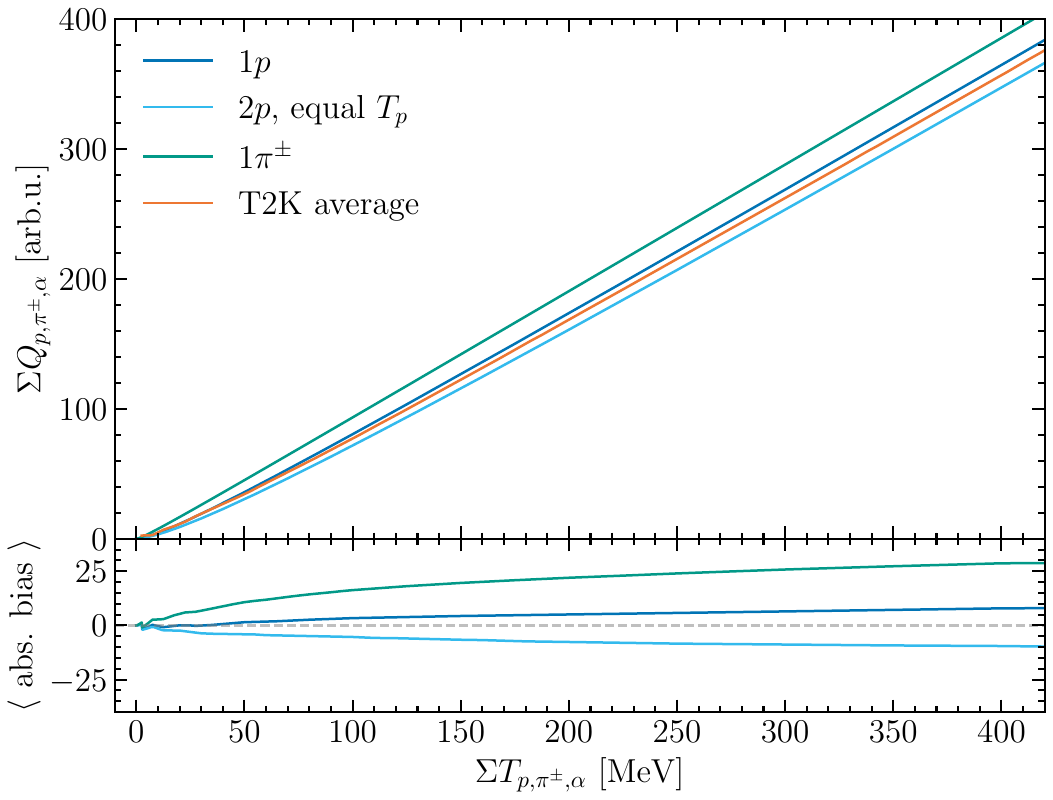}
    \caption{Correspondence between kinetic energy and visible charges of single protons and pions, as well as for combined visible charge of two protons related to the sum of their kinetic energies, assuming an even split between the two. The orange line shows the response for the T2K average particle multiplicity in CC-inclusive topologies. The right plot shows a zoom into the low end of the spectrum. The mean bias relative to the CC-inclusive average can be read off in the bottom subplots.}
    \label{fig:lookup_single_vs_avg_t2k}
\end{figure*}

Figure ~\ref{fig:lookup_single_vs_avg_t2k} shows single proton and pion, as well as two-pion responses compared to the T2K average hadronic energy response from Figure~\ref{fig:lookup_avg_refgen}.

\begin{figure*}[t]
\centering
    \includegraphics[height=0.26\textheight,clip,trim={0 0 4.7cm 0}]{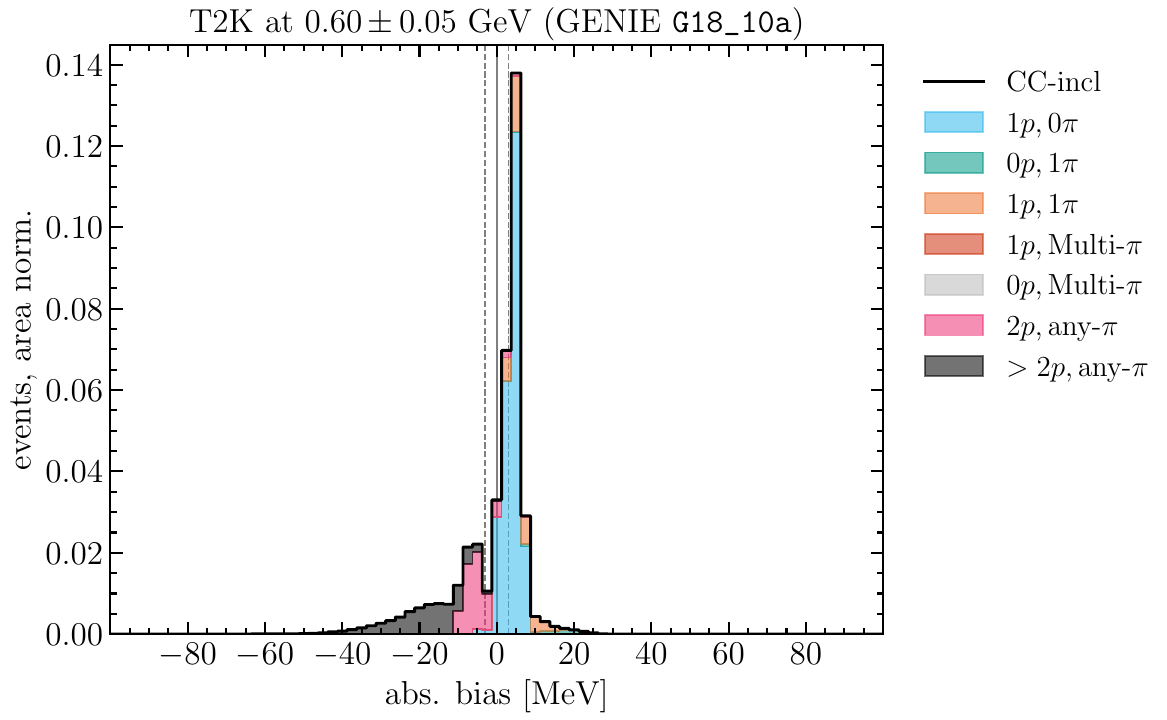}
    \includegraphics[height=0.26\textheight]{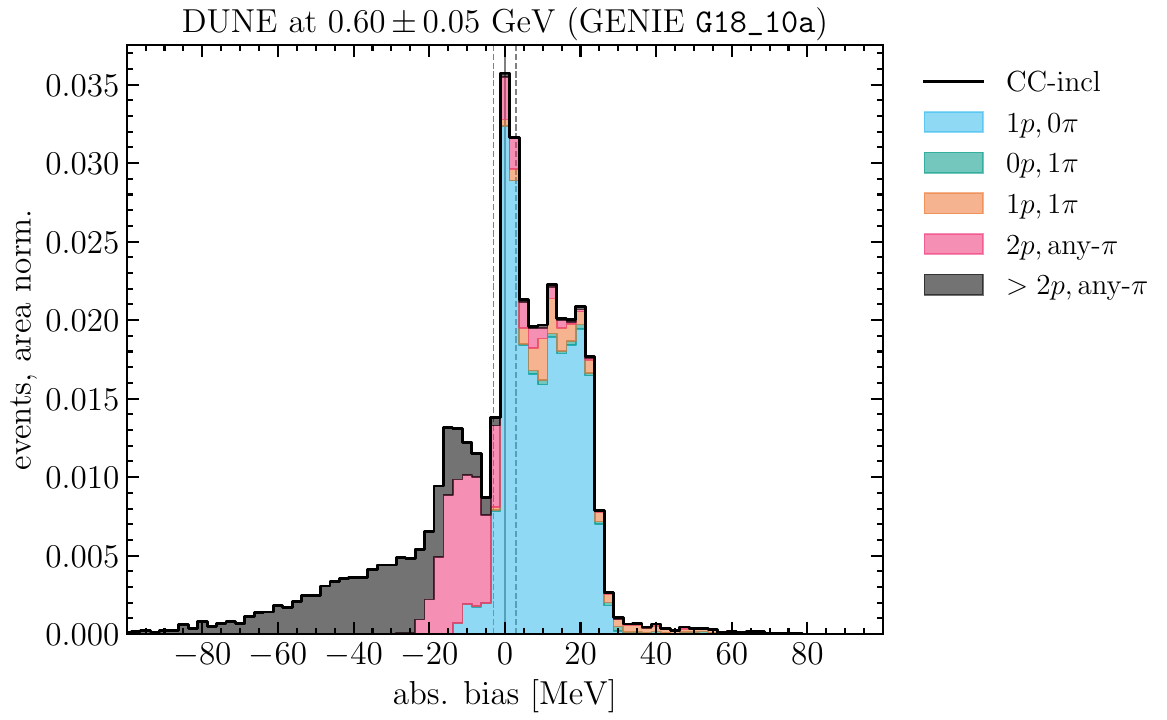}
    \caption{\label{fig:absbias_slices_bynparticle_600MeV}Bias distributions for the CC-inclusive-based calorimetric approach, showing results for $E_\nu$ slices of~(0.60$\pm$0.05)\,GeV for the T2K ND280 (left) and the DUNE FD (right) for the reference model GENIE~\texttt{G18\_10a}. Colours indicate the composition in terms of hadron multiplicities.}
\end{figure*}

Figure~\ref{fig:absbias_slices_bynparticle_600MeV} shows bias distributions for the reference model GENIE \texttt{G18\_10a} at (0.60$\pm$0.05)\,GeV for T2K and DUNE.

\begin{figure*}[p]
  \includegraphics[width=\textwidth]{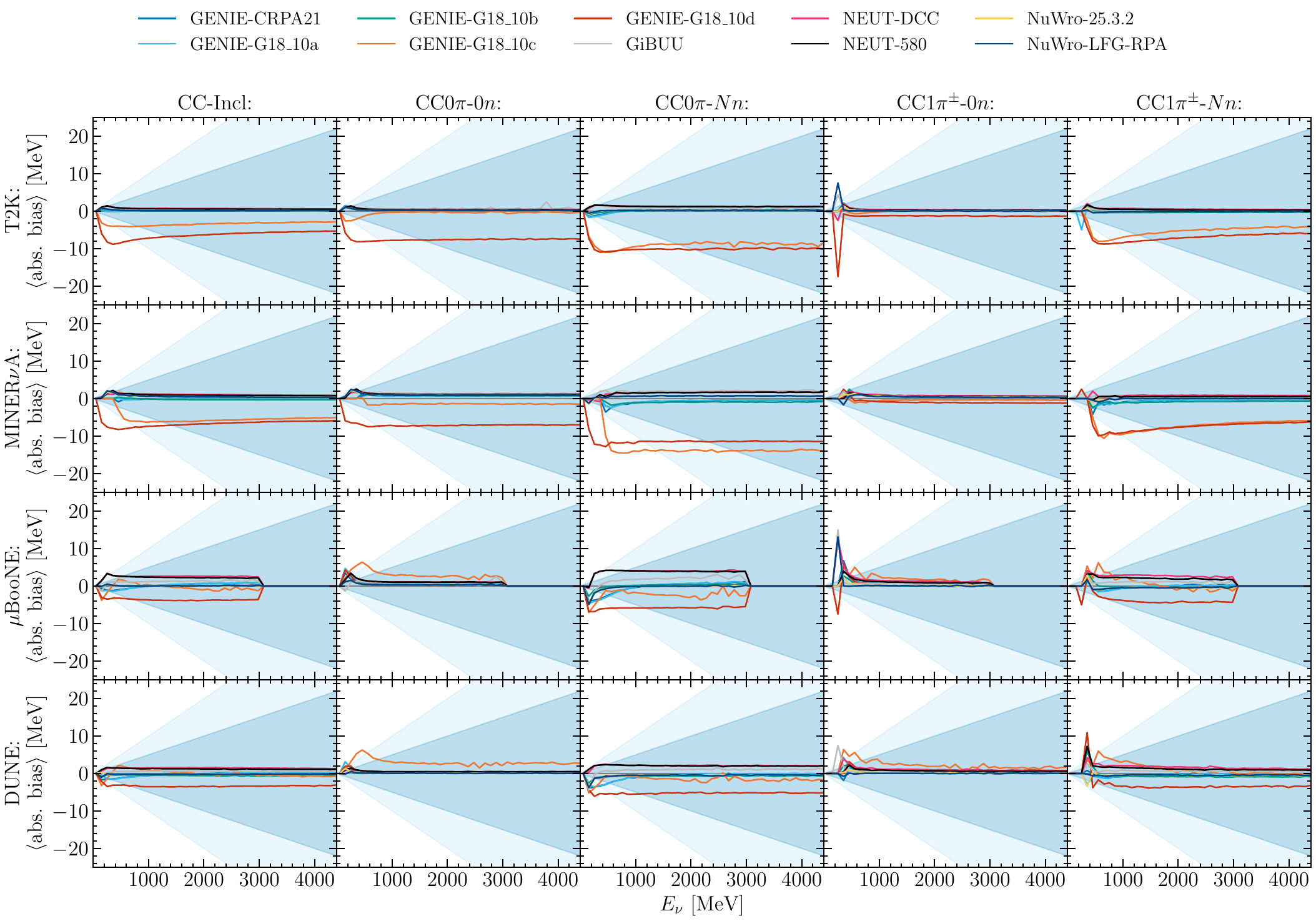}
  \caption{\label{fig:absbias_hybrid_vs_enu} 
    Mean bias for the hybrid reconstruction approach, caused only by material effects as a function for the true neutrino energy. Results are shown for T2K,  MINER$\nu$A, $\mu$BooNE, and
    DUNE (from top to bottom), with columns showing different interaction
    topologies: CC-Incl, CC$0\pi$-0n, CC$0\pi$-Nn, CC$0\pi$-Nn, CC$1\pi^\pm$-0n,
    and CC$1\pi$-Nn (from left to right).  Dark (light) shaded regions indicate $\pm$0.5\% ($\pm$1\%) bias.
    }
\end{figure*}

The model spread in the mean bias when $E_\mathrm{had,vis}$ is reconstructed using the hybrid approach (Eq.~\ref{eq:ehadvis_va_reco_hyb}) is shown as a function of $E_\nu$ in Figure~\ref{fig:absbias_hybrid_vs_enu}. Shaded regions indicate $\pm$0.5\% and $\pm$1\% of $E_\nu$. Analogous plots for other reconstruction approaches, as well as in dependence on $E_\mathrm{had,vis}$, are provided in the Supplementary Material. 

Figure~\ref{fig:boxplots_absbias_topo} shows box plots for bias distributions for a wider bias range in inclusive and exclusive topologies.

Table~\ref{tab:results_absbias} summarises mean biases for all reconstruction methods at 600\,MeV and 2.5\,GeV, with colours highlighting values exceeding $\pm$0.5\% and $\pm$1\%.

Table~\ref{tab:frac_evts_above_thresh} shows the fraction of events exceeding the target value of $\pm$0.5\% bias for each experiment and topology, and colours highlight cases where more than half of all events in the sample are affected.

\begin{table*}
    \centering
    \renewcommand{\arraystretch}{1.5}
    \setcellgapes{3pt}    
    \makegapedcells       
    \begin{tabular}{l c}
    \hline\hline
            &  FSI model\\\hline
    GENIE \texttt{G18\_10a} 
          & \texttt{hA2018I}. \cite{dytman2021comparison}  
          \\
    GENIE \texttt{G18\_10b}
          & \texttt{hN2018I}. \cite{dytman2021comparison} based on \cite{salcedo1988computer, Oset:1987re}
          \\
    GENIE \texttt{G18\_10c}
          & \citet{wright2015geant4}
          \\
    GENIE \texttt{G18\_10d}
          & \citet{cugnon2016processes,mancusi2015improving} 
          \\
    GENIE \texttt{CRPA21}
          & \texttt{hN2018I}. \cite{dytman2021comparison} based on Ref. \cite{salcedo1988computer, Oset:1987re}
          \\
    GiBUU
          & \citet{buss2012transport}  
          \\
    NuWro 25.3.2 7 
    & NuWro Cascade based on \cite{salcedo1988computer, Oset:1987re}
    \\
    NuWro LFG-RPA & NuWro Cascade based on \cite{salcedo1988computer, Oset:1987re}
    \\
    NEUT-580      
          & NEUT Cascade based on \cite{salcedo1988computer, Oset:1987re}
          \\
    NEUT-DCC      
          & NEUT Cascade based on \cite{salcedo1988computer, Oset:1987re}
          \\
       \hline\hline&
    \end{tabular}
    \caption{ \label{tab:fsi} Summary of the theoretical models to describe FSI used in this work. 
    For all cases other than GENIE \texttt{G18\_10d} the simulation samples provided in Ref.~\cite{dolan2026cpviolation} were used, which provides additional details on the theoretical models.}
\vspace{4em}
    \begin{tabular}{l c c c c c}
    \hline\hline
            &  Nucl. GS & QES & 2p2h & RES & SIS/DIS\\\hline
    GENIE \texttt{G18\_10a} & LFG & \citet{Nieves:2011pp} 
                  & \citet{Nieves:2011pp} 
                  & \citet{Berger:2007rq}    
                  &
                  \makecell{\citet{Bodek:2003wc}\\\citet{Yang:2009zx}}
                  \\
    GENIE \texttt{G18\_10b} & LFG& \citet{Nieves:2011pp} 
                  & \citet{Nieves:2011pp} 
                  & \citet{Berger:2007rq}    
                  &
                  \makecell{\citet{Bodek:2003wc}\\\citet{Yang:2009zx}}
                  \\
    GENIE \texttt{G18\_10c} & LFG& \citet{Nieves:2011pp} 
                  & \citet{Nieves:2011pp} 
                  & \citet{Berger:2007rq}    
                  &
                  \makecell{\citet{Bodek:2003wc}\\\citet{Yang:2009zx}}
                  \\
    GENIE \texttt{G18\_10d} & LFG& \citet{Nieves:2011pp} 
                  & \citet{Nieves:2011pp} 
                  & \citet{Berger:2007rq}    
                  &
                  \makecell{\citet{Bodek:2003wc}\\\citet{Yang:2009zx}}
                  \\
    GENIE \texttt{CRPA21} 
          & \makecell{LFG,\,$q_3$-dep. \\$E_\mathrm{rmv}$~\cite{Dolan:2019bxf}} 
          & \citet{Jachowicz:2002rr} 
          & \citet{RuizSimo:2016ikw} 
          & \citet{Berger:2007rq}
          & \makecell{\citet{Bodek:2003wc}\\\citet{Yang:2009zx}}
          
          \\
    GiBUU
          & LFG 
          & \citet{Leitner:2008ue} 
          & \citet{Gallmeister:2016dnq} 
          & Lalakulich~\textit{et al.} \cite{Lalakulich:2012cj,Lalakulich:2010ss} 
          & \citet{Mosel:2023zek}
          \\
    NuWro 25.3.2 7 
    & LFG 
    & \citet{Graczyk:2003ru} 
    & \citet{Nieves:2011pp}  
    &  \citet{Adler:1975mt}  
    & \citet{Adler:1975mt}
    \\
    NuWro LFG-RPA 
    & LFG 
    & \citet{Graczyk:2003ru} 
    & \makecell{\citet{Sobczyk:2020dkn} \\ \citet{Prasad:2024gnv}} 
    & \citet{Yan:2024kkg} 
    & \citet{Yan:2024kkg}
    \\\
    NEUT-580      
          & LFG/RFG 
          & \citet{Nieves:2011pp} 
          & \citet{Nieves:2011pp}  
          & \citet{Berger:2007rq} 
          & \makecell{\citet{Bodek:2003wc}\\\citet{NuSTEC:2020nsl}}
          \\
    NEUT-DCC      
          & LFG/RFG 
          & \citet{Nieves:2011pp} 
          & \citet{Nieves:2011pp}  
          & \citet{Nakamura:2015rta} 
          & \makecell{\citet{Bodek:2003wc}\\\citet{NuSTEC:2020nsl}}
          \\
       \hline\hline
    \end{tabular}
    \caption{ \label{tab:models} Summary of the theoretical models to describe the
    nuclear ground state, CC-QES, 2p2h, RES and SIS/DIS interactions used in this work. 
    For all cases other than GENIE \texttt{G18\_10d} the simulation samples provided in Ref.~\cite{dolan2026cpviolation} were used, which provides additional details on the theoretical models.}
\end{table*}

\begin{figure*}[p]
    \begin{subfigure}{0.49\textwidth}
        \includegraphics[height=0.27\textheight,trim={0 0 0 0.5cm},clip]{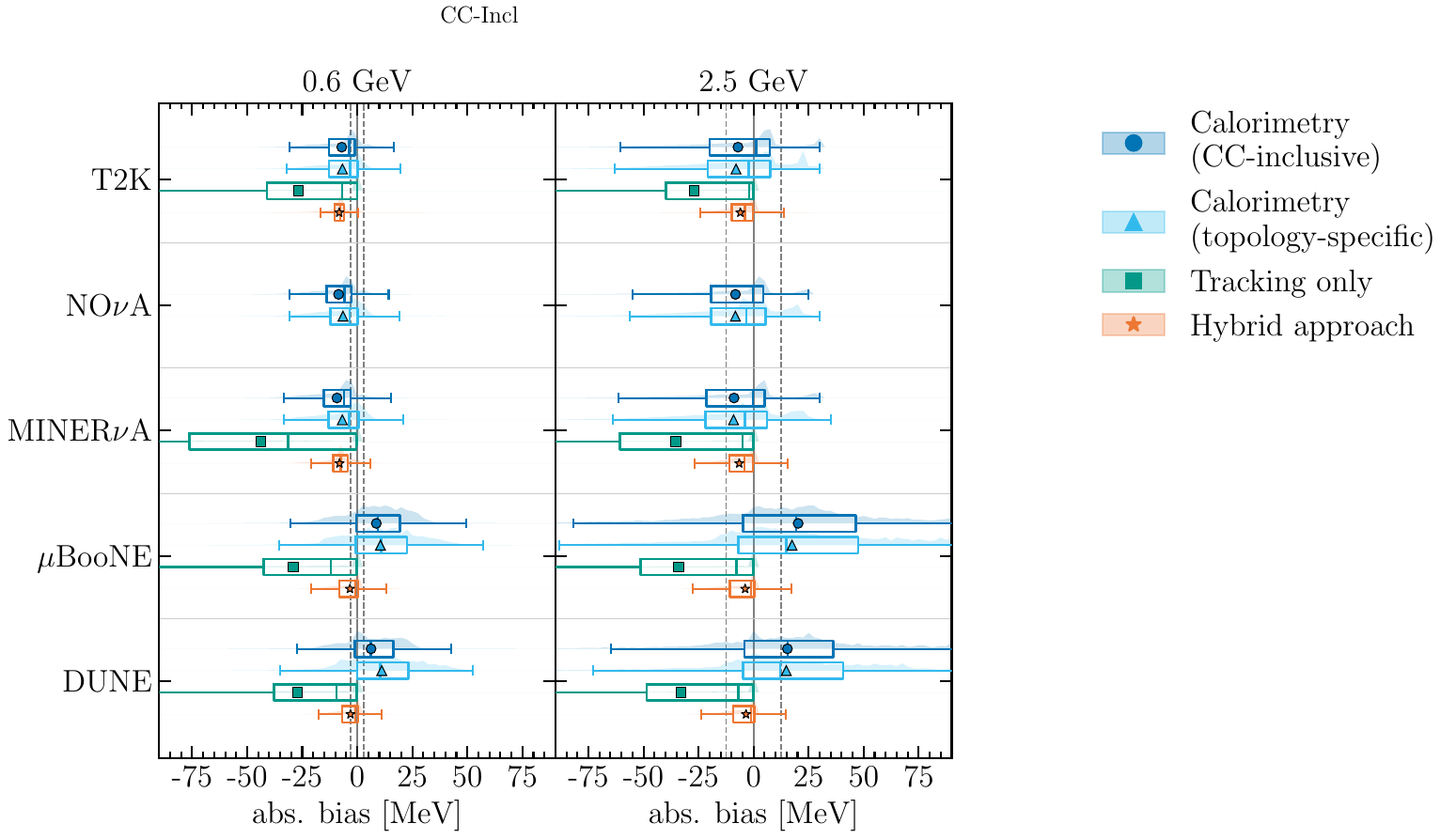}
        \caption{CC-inclusive}
    \end{subfigure}
    
    \begin{subfigure}{0.49\textwidth}
        \includegraphics[height=0.27\textheight,trim={0 0 9cm 0.5cm},clip]{figures/bias_ranges_bias_Enu_fromquench_CCIncl_v6.pdf}
        \caption{CC$0\pi$-$0n$}
    \end{subfigure}
    \begin{subfigure}{0.49\textwidth}
        \includegraphics[height=0.27\textheight,trim={0 0 9cm 0.5cm},clip]{figures/bias_ranges_bias_Enu_fromquench_CCIncl_v6.pdf}
        \caption{CC$0\pi$-$Nn$}
    \end{subfigure}
    \begin{subfigure}{0.49\textwidth}
        \includegraphics[height=0.27\textheight,trim={0 0 9cm 0.5cm},clip]{figures/bias_ranges_bias_Enu_fromquench_CCIncl_v6.pdf}
        \caption{CC$1\pi^\pm 0n$}
    \end{subfigure}
    \begin{subfigure}{0.49\textwidth}
        \includegraphics[height=0.27\textheight,trim={0 0 9cm 0.5cm},clip]{figures/bias_ranges_bias_Enu_fromquench_CCIncl_v6.pdf}
        \caption{CC$1\pi^\pm Nn$}
    \end{subfigure}

    \caption{Mean bias caused by material effects for slices of $E_\nu^{true}$ of (0.60$\pm$0.05)\,GeV and (2.50$\pm$0.05)\,GeV in exclusive interaction topologies, showing mean absolute bias (dot), box plots indicating the central 50\% of events (interquartile range, IQR) with the median as vertical line and whiskers to $1.5\times$IQR. Shaded histograms show the full bias distribution. Vertical dashed lines indicate $\pm$0.5\% of the given value of $E_\nu$.}
    \label{fig:boxplots_absbias_topo}
\end{figure*}

\begin{table*}[p]
\centering

\vspace{1em}
Calorimetric method with CC-inclusive-based reconstruction:\phantom{yg}

\begin{tabular}{c|c c|c c|c c|c c|c c}
\hline\hline
 & \multicolumn{2}{c|}{CC-Incl} & \multicolumn{2}{c|}{CC$0\pi$-$0n$} & \multicolumn{2}{c|}{CC$0\pi$-$Nn$} & \multicolumn{2}{c|}{CC$1\pi^\pm$-$0n$} & \multicolumn{2}{c}{CC$1\pi^\pm$-$Nn$} \\
 & 0.6\,GeV & 2.5\,GeV & 0.6\,GeV & 2.5\,GeV & 0.6\,GeV & 2.5\,GeV & 0.6\,GeV & 2.5\,GeV & 0.6\,GeV & 2.5\,GeV \\\hline
T2K& \colorbox{highlight2}{-7.1}& -7.1& \colorbox{highlight1}{-5.0}& -4.0& \colorbox{highlight2}{-19.7}& \colorbox{highlight2}{-31.7}& \colorbox{highlight2}{8.8}& 5.8& \colorbox{highlight2}{13.8}& \colorbox{highlight1}{16.0}\\
NO$\nu$A& \colorbox{highlight2}{-8.5}& -8.2& \colorbox{highlight2}{-6.8}& -6.5& \colorbox{highlight2}{-21.4}& \colorbox{highlight2}{-31.3}& \colorbox{highlight1}{5.9}& 2.8& \colorbox{highlight2}{9.7}& 11.9\\
MINER$\nu$A& \colorbox{highlight2}{-9.2}& -8.9& \colorbox{highlight2}{-7.3}& -6.4& \colorbox{highlight2}{-21.3}& \colorbox{highlight2}{-33.6}& \colorbox{highlight2}{6.5}& 3.7& \colorbox{highlight2}{10.9}& \colorbox{highlight1}{14.1}\\
$\mu$BooNE& \colorbox{highlight2}{8.7}& \colorbox{highlight1}{20.3}& \colorbox{highlight2}{14.9}& \colorbox{highlight2}{28.5}& \colorbox{highlight2}{-20.0}& \colorbox{highlight2}{-46.7}& \colorbox{highlight2}{33.4}& \colorbox{highlight2}{37.8}& \colorbox{highlight2}{39.3}& \colorbox{highlight2}{48.4}\\
DUNE& \colorbox{highlight2}{6.3}& \colorbox{highlight1}{15.4}& \colorbox{highlight2}{12.3}& \colorbox{highlight1}{22.6}& \colorbox{highlight2}{-19.0}& \colorbox{highlight2}{-44.5}& \colorbox{highlight2}{25.8}& \colorbox{highlight2}{31.1}& \colorbox{highlight2}{30.9}& \colorbox{highlight2}{40.2}\\
\hline\hline
\end{tabular}

\vspace{1em}
Calorimetric method with topology-specific reconstruction:\phantom{yg}

\begin{tabular}{c|c c|c c|c c|c c|c c}
\hline\hline
 & \multicolumn{2}{c|}{CC-Incl} & \multicolumn{2}{c|}{CC$0\pi$-$0n$} & \multicolumn{2}{c|}{CC$0\pi$-$Nn$} & \multicolumn{2}{c|}{CC$1\pi^\pm$-$0n$} & \multicolumn{2}{c}{CC$1\pi^\pm$-$Nn$} \\
 & 0.6\,GeV & 2.5\,GeV & 0.6\,GeV & 2.5\,GeV & 0.6\,GeV & 2.5\,GeV & 0.6\,GeV & 2.5\,GeV & 0.6\,GeV & 2.5\,GeV \\\hline
T2K& \colorbox{highlight2}{-6.7}& -8.0& \colorbox{highlight1}{4.7}& 8.1& \colorbox{highlight2}{-18.3}& \colorbox{highlight2}{-26.3}& -2.6& -4.2& \colorbox{highlight2}{-14.9}& 8.6\\
NO$\nu$A& \colorbox{highlight2}{-6.5}& -8.3& \colorbox{highlight1}{5.0}& 7.8& \colorbox{highlight2}{-18.1}& \colorbox{highlight1}{-23.7}& \colorbox{highlight1}{-4.4}& -4.9& \colorbox{highlight2}{-14.9}& -8.6\\
MINER$\nu$A& \colorbox{highlight2}{-6.8}& -9.1& \colorbox{highlight1}{5.5}& 8.7& \colorbox{highlight2}{-17.6}& \colorbox{highlight2}{-25.2}& \colorbox{highlight2}{-6.3}& -6.3& \colorbox{highlight2}{-15.9}& -9.4\\
$\mu$BooNE& \colorbox{highlight2}{10.5}& \colorbox{highlight1}{17.5}& \colorbox{highlight2}{20.1}& \colorbox{highlight2}{38.4}& \colorbox{highlight2}{-16.2}& \colorbox{highlight2}{-39.2}& \colorbox{highlight2}{-8.8}& \colorbox{highlight1}{14.5}& \colorbox{highlight2}{-14.0}& \colorbox{highlight1}{24.6}\\
DUNE& \colorbox{highlight2}{11.1}& \colorbox{highlight1}{14.9}& \colorbox{highlight2}{21.6}& \colorbox{highlight2}{41.6}& \colorbox{highlight2}{-12.4}& \colorbox{highlight2}{-29.7}& \colorbox{highlight2}{-12.0}& -8.0& \colorbox{highlight2}{-16.3}& \colorbox{highlight2}{-26.8}\\
\hline\hline
\end{tabular}

\vspace{1em}
Tracking-based reconstruction:\phantom{yg}

\begin{tabular}{c|c c|c c|c c|c c|c c}
\hline\hline
 & \multicolumn{2}{c|}{CC-Incl} & \multicolumn{2}{c|}{CC$0\pi$-$0n$} & \multicolumn{2}{c|}{CC$0\pi$-$Nn$} & \multicolumn{2}{c|}{CC$1\pi^\pm$-$0n$} & \multicolumn{2}{c}{CC$1\pi^\pm$-$Nn$} \\
 & 0.6\,GeV & 2.5\,GeV & 0.6\,GeV & 2.5\,GeV & 0.6\,GeV & 2.5\,GeV & 0.6\,GeV & 2.5\,GeV & 0.6\,GeV & 2.5\,GeV \\\hline
T2K& \colorbox{highlight2}{-26.7}& \colorbox{highlight2}{-27.1}& \colorbox{highlight2}{-17.2}& \colorbox{highlight1}{-14.0}& \colorbox{highlight2}{-48.3}& \colorbox{highlight2}{-58.9}& \colorbox{highlight2}{-15.4}& -8.0& \colorbox{highlight2}{-40.6}& \colorbox{highlight2}{-29.1}\\
NO$\nu$A& \colorbox{highlight2}{-44.1}& \colorbox{highlight2}{-33.9}& \colorbox{highlight2}{-32.4}& \colorbox{highlight2}{-28.5}& \colorbox{highlight2}{-69.9}& \colorbox{highlight2}{-75.7}& \colorbox{highlight2}{-36.9}& \colorbox{highlight1}{-15.3}& \colorbox{highlight2}{-55.5}& \colorbox{highlight2}{-37.1}\\
MINER$\nu$A& \colorbox{highlight2}{-43.8}& \colorbox{highlight2}{-35.4}& \colorbox{highlight2}{-33.4}& \colorbox{highlight2}{-27.7}& \colorbox{highlight2}{-68.3}& \colorbox{highlight2}{-75.4}& \colorbox{highlight2}{-36.2}& \colorbox{highlight1}{-14.3}& \colorbox{highlight2}{-52.8}& \colorbox{highlight2}{-36.7}\\
$\mu$BooNE& \colorbox{highlight2}{-29.1}& \colorbox{highlight2}{-34.0}& \colorbox{highlight2}{-12.0}& -7.3& \colorbox{highlight2}{-39.6}& \colorbox{highlight2}{-52.7}& \colorbox{highlight2}{-11.8}& -5.8& \colorbox{highlight2}{-37.6}& \colorbox{highlight2}{-34.9}\\
DUNE& \colorbox{highlight2}{-27.1}& \colorbox{highlight2}{-33.0}& \colorbox{highlight2}{-10.9}& -6.9& \colorbox{highlight2}{-37.2}& \colorbox{highlight2}{-51.0}& \colorbox{highlight2}{-10.8}& -3.4& \colorbox{highlight2}{-34.7}& \colorbox{highlight2}{-34.1}\\
\hline\hline
\end{tabular}

\vspace{1em}
Hybrid method with topology-specific reconstruction:\phantom{g}

\begin{tabular}{c|c c|c c|c c|c c|c c}
\hline\hline
 & \multicolumn{2}{c|}{CC-Incl} & \multicolumn{2}{c|}{CC$0\pi$-$0n$} & \multicolumn{2}{c|}{CC$0\pi$-$Nn$} & \multicolumn{2}{c|}{CC$1\pi^\pm$-$0n$} & \multicolumn{2}{c}{CC$1\pi^\pm$-$Nn$} \\
 & 0.6\,GeV & 2.5\,GeV & 0.6\,GeV & 2.5\,GeV & 0.6\,GeV & 2.5\,GeV & 0.6\,GeV & 2.5\,GeV & 0.6\,GeV & 2.5\,GeV \\\hline
T2K& \colorbox{highlight2}{-8.2}& -6.0& \colorbox{highlight2}{-8.1}& -7.5& \colorbox{highlight2}{-10.9}& -9.9& -1.3& -1.2& \colorbox{highlight2}{-8.7}& -6.8\\
MINER$\nu$A& \colorbox{highlight2}{-8.1}& -6.5& \colorbox{highlight2}{-7.5}& -7.1& \colorbox{highlight2}{-13.9}& \colorbox{highlight1}{-14.3}& 1.6& -1.1& \colorbox{highlight2}{-10.0}& -7.2\\
$\mu$BooNE& \colorbox{highlight1}{-3.3}& -3.8& \colorbox{highlight1}{5.4}& 2.6& \colorbox{highlight1}{-5.9}& -6.1& \colorbox{highlight1}{5.0}& 1.1& \colorbox{highlight2}{6.2}& -4.3\\
DUNE& \colorbox{highlight1}{-3.1}& -3.4& \colorbox{highlight1}{5.4}& 2.9& \colorbox{highlight1}{-5.4}& -5.0& \colorbox{highlight1}{5.6}& 1.2& \colorbox{highlight1}{6.0}& -3.7\\
\hline\hline
\end{tabular}
\caption{\label{tab:results_absbias}Model spread in mean absolute bias (in MeV) across all experiments and all reconstruction methods considered in this study, evaluated at the two reference values of $E_\nu=$600\,MeV and 2.5\,GeV. Values exceeding 0.5\% of the neutrino energy value are highlighted in blue, and those exceeding 1\% are highlighted in orange.}
\end{table*}

\begin{table*}[p]
\centering

\vspace{1em}
Calorimetric method with CC-inclusive-based reconstruction:\phantom{yg}

\begin{tabular}{c|c c|c c|c c|c c|c c}
\hline\hline
 & \multicolumn{2}{c|}{CC-Incl} & \multicolumn{2}{c|}{CC$0\pi$-$0n$} & \multicolumn{2}{c|}{CC$0\pi$-$Nn$} & \multicolumn{2}{c|}{CC$1\pi^\pm$-$0n$} & \multicolumn{2}{c}{CC$1\pi^\pm$-$Nn$} \\
 & 0.6\,GeV & 2.5\,GeV & 0.6\,GeV & 2.5\,GeV & 0.6\,GeV & 2.5\,GeV & 0.6\,GeV & 2.5\,GeV & 0.6\,GeV & 2.5\,GeV \\\hline
T2K& \colorbox{highlight1}{79.7}& 45.1& \colorbox{highlight1}{78.2}& 13.9& \colorbox{highlight1}{94.7}& \colorbox{highlight1}{74.9}& \colorbox{highlight1}{88.8}& 15.4& \colorbox{highlight1}{94.1}& \colorbox{highlight1}{72.5}\\
NO$\nu$A& \colorbox{highlight1}{74.8}& 43.3& \colorbox{highlight1}{70.8}& 15.9& \colorbox{highlight1}{97.2}& \colorbox{highlight1}{76.2}& \colorbox{highlight1}{57.4}& 11.5& \colorbox{highlight1}{83.3}& \colorbox{highlight1}{71.8}\\
MINER$\nu$A& \colorbox{highlight1}{75.1}& 46.3& \colorbox{highlight1}{70.6}& 16.9& \colorbox{highlight1}{97.4}& \colorbox{highlight1}{76.6}& \colorbox{highlight1}{64.3}& 14.6& \colorbox{highlight1}{84.8}& \colorbox{highlight1}{73.2}\\
$\mu$BooNE& \colorbox{highlight1}{88.4}& \colorbox{highlight1}{80.9}& \colorbox{highlight1}{93.5}& \colorbox{highlight1}{78.4}& \colorbox{highlight1}{84.2}& \colorbox{highlight1}{75.0}& \colorbox{highlight1}{99.1}& \colorbox{highlight1}{92.7}& \colorbox{highlight1}{97.8}& \colorbox{highlight1}{90.6}\\
DUNE& \colorbox{highlight1}{83.7}& \colorbox{highlight1}{78.0}& \colorbox{highlight1}{87.1}& \colorbox{highlight1}{73.8}& \colorbox{highlight1}{81.4}& \colorbox{highlight1}{73.2}& \colorbox{highlight1}{98.2}& \colorbox{highlight1}{93.6}& \colorbox{highlight1}{96.6}& \colorbox{highlight1}{87.9}\\
\hline\hline
\end{tabular}

\vspace{1em}
Calorimetric method with topology-specific reconstruction:\phantom{yg}

\begin{tabular}{c|c c|c c|c c|c c|c c}
\hline\hline
 & \multicolumn{2}{c|}{CC-Incl} & \multicolumn{2}{c|}{CC$0\pi$-$0n$} & \multicolumn{2}{c|}{CC$0\pi$-$Nn$} & \multicolumn{2}{c|}{CC$1\pi^\pm$-$0n$} & \multicolumn{2}{c}{CC$1\pi^\pm$-$Nn$} \\
 & 0.6\,GeV & 2.5\,GeV & 0.6\,GeV & 2.5\,GeV & 0.6\,GeV & 2.5\,GeV & 0.6\,GeV & 2.5\,GeV & 0.6\,GeV & 2.5\,GeV \\\hline
T2K& \colorbox{highlight1}{80.9}& 49.1& \colorbox{highlight1}{86.4}& 37.1& \colorbox{highlight1}{91.8}& \colorbox{highlight1}{69.6}& 42.1& 16.5& \colorbox{highlight1}{83.1}& \colorbox{highlight1}{75.0}\\
NO$\nu$A& \colorbox{highlight1}{78.3}& 46.1& \colorbox{highlight1}{79.3}& 34.9& \colorbox{highlight1}{90.8}& \colorbox{highlight1}{67.2}& \colorbox{highlight1}{52.2}& 14.6& \colorbox{highlight1}{80.6}& \colorbox{highlight1}{74.2}\\
MINER$\nu$A& \colorbox{highlight1}{81.7}& \colorbox{highlight1}{50.4}& \colorbox{highlight1}{83.0}& 42.7& \colorbox{highlight1}{91.0}& \colorbox{highlight1}{68.2}& \colorbox{highlight1}{69.5}& 16.9& \colorbox{highlight1}{84.1}& \colorbox{highlight1}{76.6}\\
$\mu$BooNE& \colorbox{highlight1}{91.2}& \colorbox{highlight1}{76.4}& \colorbox{highlight1}{96.7}& \colorbox{highlight1}{82.8}& \colorbox{highlight1}{88.3}& \colorbox{highlight1}{72.4}& \colorbox{highlight1}{89.5}& \colorbox{highlight1}{52.7}& \colorbox{highlight1}{89.1}& \colorbox{highlight1}{84.3}\\
DUNE& \colorbox{highlight1}{86.7}& \colorbox{highlight1}{72.0}& \colorbox{highlight1}{94.2}& \colorbox{highlight1}{80.7}& \colorbox{highlight1}{82.1}& \colorbox{highlight1}{68.3}& \colorbox{highlight1}{90.6}& 40.0& \colorbox{highlight1}{88.1}& \colorbox{highlight1}{82.9}\\
\hline\hline
\end{tabular}

\vspace{1em}
Tracking-based reconstruction:\phantom{yg}

\begin{tabular}{c|c c|c c|c c|c c|c c}
\hline\hline
 & \multicolumn{2}{c|}{CC-Incl} & \multicolumn{2}{c|}{CC$0\pi$-$0n$} & \multicolumn{2}{c|}{CC$0\pi$-$Nn$} & \multicolumn{2}{c|}{CC$1\pi^\pm$-$0n$} & \multicolumn{2}{c}{CC$1\pi^\pm$-$Nn$} \\
 & 0.6\,GeV & 2.5\,GeV & 0.6\,GeV & 2.5\,GeV & 0.6\,GeV & 2.5\,GeV & 0.6\,GeV & 2.5\,GeV & 0.6\,GeV & 2.5\,GeV \\\hline
T2K& \colorbox{highlight1}{90.5}& 40.6& \colorbox{highlight1}{99.2}& 24.5& \colorbox{highlight1}{98.7}& \colorbox{highlight1}{87.8}& 34.5& 13.8& \colorbox{highlight1}{83.4}& 45.9\\
NO$\nu$A& \colorbox{highlight1}{91.1}& 43.2& \colorbox{highlight1}{99.5}& 43.6& \colorbox{highlight1}{99.2}& \colorbox{highlight1}{91.1}& \colorbox{highlight1}{59.3}& 24.2& \colorbox{highlight1}{84.3}& 48.6\\
MINER$\nu$A& \colorbox{highlight1}{93.4}& 44.9& \colorbox{highlight1}{99.7}& 40.6& \colorbox{highlight1}{99.1}& \colorbox{highlight1}{91.1}& \colorbox{highlight1}{61.3}& 22.4& \colorbox{highlight1}{84.7}& 49.0\\
$\mu$BooNE& \colorbox{highlight1}{58.7}& 46.2& 32.6& 19.5& \colorbox{highlight1}{85.9}& \colorbox{highlight1}{70.4}& 33.4& 15.5& \colorbox{highlight1}{74.2}& 45.5\\
DUNE& \colorbox{highlight1}{56.8}& 45.0& 25.6& 22.4& \colorbox{highlight1}{85.2}& \colorbox{highlight1}{67.2}& 30.4& 10.9& \colorbox{highlight1}{72.8}& 45.0\\
\hline\hline
\end{tabular}

\vspace{1em}
Hybrid method with topology-specific reconstruction:\phantom{g}

\begin{tabular}{c|c c|c c|c c|c c|c c}
\hline\hline
 & \multicolumn{2}{c|}{CC-Incl} & \multicolumn{2}{c|}{CC$0\pi$-$0n$} & \multicolumn{2}{c|}{CC$0\pi$-$Nn$} & \multicolumn{2}{c|}{CC$1\pi^\pm$-$0n$} & \multicolumn{2}{c}{CC$1\pi^\pm$-$Nn$} \\
 & 0.6\,GeV & 2.5\,GeV & 0.6\,GeV & 2.5\,GeV & 0.6\,GeV & 2.5\,GeV & 0.6\,GeV & 2.5\,GeV & 0.6\,GeV & 2.5\,GeV \\\hline
T2K& \colorbox{highlight1}{83.9}& 18.0& \colorbox{highlight1}{95.4}& 10.3& \colorbox{highlight1}{87.9}& 35.7& 22.3& 5.7& \colorbox{highlight1}{70.1}& 20.7\\
MINER$\nu$A& \colorbox{highlight1}{81.7}& 20.7& \colorbox{highlight1}{90.9}& 10.9& \colorbox{highlight1}{89.4}& 44.0& 24.6& 4.9& \colorbox{highlight1}{72.2}& 23.1\\
$\mu$BooNE& 48.2& 31.4& 17.5& 7.2& \colorbox{highlight1}{73.8}& 46.4& 26.8& 5.7& \colorbox{highlight1}{65.3}& 32.3\\
DUNE& 43.3& 28.1& 12.0& 4.0& \colorbox{highlight1}{66.7}& 41.1& 20.2& 2.4& \colorbox{highlight1}{60.0}& 30.5\\
\hline\hline
\end{tabular}
\caption{\label{tab:frac_evts_above_thresh}Fraction of events exceeding 0.5\% absolute bias at $E_
u$ of 0.6$\pm$0.1\,GeV and 2.5$\pm$0.1\,GeV, for different methods in MeV. Colours highlight scenarios where more than half of the events exceed this bias threshold.}
\end{table*}

\end{document}